\documentclass[10pt]{elsarticle}

\usepackage[margin=1.0in]{geometry}
\usepackage{algorithm}
\usepackage{algorithmicx}
\usepackage[noend]{algpseudocode}
\usepackage{graphicx}
\usepackage{tikz}
\usetikzlibrary{automata, arrows.meta, positioning}
\usepackage{subcaption}
\usepackage{caption}
\usepackage{amsmath}

\usepackage{multirow}
\usepackage{array}
\usepackage{tabularx}
\usepackage{color,soul}
\newcolumntype{L}[1]{>{\raggedright\let\newline\\\arraybackslash\hspace{0pt}}m{#1}}
\newcolumntype{C}[1]{>{\centering\let\newline\\\arraybackslash\hspace{0pt}}m{#1}}
\newcolumntype{R}[1]{>{\raggedleft\let\newline\\\arraybackslash\hspace{0pt}}m{#1}}

\algdef{SE}[DOWHILE]{Do}{doWhile}{\algorithmicdo}[1]{\algorithmicwhile\ #1}%

\let\today\relax
\makeatletter
\def\ps@pprintTitle{%
    \let\@oddhead\@empty
    \let\@evenhead\@empty
    \def\@oddfoot{\footnotesize\itshape
         {} \hfill\today}%
    \let\@evenfoot\@oddfoot
    }
\makeatother

\begin{document}

\begin{frontmatter}

\title{Regulating CPU Temperature With Thermal-Aware Scheduling Using a Reduced Order Learning Thermal Model}

\author{Anthony Dowling}

\author{Lin Jiang}
\author{Ming-Cheng Cheng}
\author{Yu Liu}

    \address{Department of Electrical and Computer Engineering, Clarkson University, 8 Clarkson Ave., Potsdam, 13699, New York, USA}
\begin{abstract}
    
    Modern real-time systems utilize considerable amounts of
    power while executing computation-intensive tasks. The
    execution of these tasks leads to significant power
    dissipation and heating of the device. It therefore
    results in severe thermal issues like temperature
    escalation, high thermal gradients, and excessive hot
    spot formation, which may result in degrading chip
    performance, accelerating device aging, and premature
    failure.  Thermal-Aware Scheduling (TAS) enables
    optimization of thermal dissipation to maintain a safe
    thermal state.  In this work, we implement a new TAS
    algorithm, POD-TAS, which manages the thermal behavior
    of a multi-core CPU based on a defined set of states and their
    transitions.  We compare the performances of a dynamic
    RC thermal circuit simulator
    (HotSpot~\cite{hotspot_2003}) and a reduced order Proper
    Orthogonal Decomposition (POD)-based thermal model and
    we select the latter for use in our POD-TAS algorithm.
    We implement a novel simulation-based evaluation
    methodology to compare TAS algorithms. This methodology
    is used to evaluate the performance of the proposed
    POD-TAS algorithm.  Additionally, we compare the
    performance of a state of the art TAS algorithm,
    RT-TAS~\cite{lee_2019}, to our proposed POD-TAS
    algorithm.  Furthermore, we utilize the COMBS benchmark
    suite~\cite{combs_2020} to provide CPU workloads for
    task scheduling. Our experimental results on a
    multi-core processor using a set of 4 benchmarks
    demonstrate that the proposed POD-TAS method can improve
    thermal performance by decreasing the peak thermal
    variance by 53.0\% and the peak chip temperature of
    29.01\%.  Using a set of 8 benchmarks, the comparison of
    the two algorithms shows a decrease of 29.57\% in the
    peak spatial variance of the chip temperature and
    26.26\% in the peak chip temperature. We also identify
    several potential future research directions.

\end{abstract}


\end{frontmatter}

\section{Introduction}
\label{sec:intro}

Real-time computing systems aim to execute computational tasks to meet
real-world clock timing deadlines as required by their use case.  Modern
embedded computing chips used in real-time systems are comprised of large
numbers of transistors which execute computationally massive tasks.  However,
these complex devices suffer from large amounts of heat dissipation due to the
electricity required to power the integrated circuits.  This heat dissipation
can create a number of issues within the chip's circuitry, including mechanical
stress and electrical malfunction.  These issues can cause the chip to be unable
to meet the real-time deadlines of tasks. This has led to the implementation of
methods that can help to minimize the heat dissipation to avoid hardware damage
and maintain the ability to meet real-time deadlines for tasks. These
improvements can aid in increasing chip reliablity for safety critical systems
such as autonomous vehicles, in-flight computers, and others.  This, in turn,
improves the level of safety that these systems can maintain for their
operators.

Various measures have been proposed to control the thermal behavior of modern
embedded chips to address the issue of thermal stress. These measures include
both reactive and proactive methods~\cite{krishna_2017}.  A reactive method
takes action when a temperature becomes too high, while a proactive method aims
to avoid the temperature rising too high to begin with.  Reactive methods
include clock gating, power gating, and Dynamic Voltage and Frequency Scaling
(DVFS).  Of these, DVFS is a popular solution~\cite{krishna_2017,zhou_2020}, but
since it reduces power consumption by lowering the chip voltage and operating
frequency, it results in degradation of the performance, which may become an
obstacle to time-sensitive real-time systems~\cite{xu_2019}.  The major argument
in favor of using DVFS is that while performance is sacrificed, deadlines can
still be met~\cite{krishna_2017}. The success of this argument relies on a
moderate and predictable workload so that the scheduler can generate enough
slack time while still maintaining the thermal state of the chip.  A heavy or
sporadic workload challenges DVFS and requires a more effective scheduling
approach to maintain real-time task deadlines. Furthermore, not all available
chips support the use of DVFS, due to an incapability to modify the chip
operating frequency. This limits the generalizability of algorithms that rely on
DVFS.

To address the thermal stress issues proactively without negatively impacting
the performance of the multi-core processor, thermal-aware task scheduling aims
to optimize heat dissipation while maintaining the deadlines of tasks as
required by hard real-time systems. By applying Thermal-Aware Scheduling (TAS),
the thermal output is lowered, which minimizes the cooling costs, energy
consumption, and hazards to the chip's reliability. Thus, TAS is a very active
research area~\cite{krishna_2017,zhou_2020}.  However, to enhance the
effectiveness of TAS, accurate and efficient thermal simulation of the CPU is
required. In the past, many methods for thermal simulation have been developed,
and each carries differing levels of accuracy and efficiency.

Rigorous approaches to thermal simulation that provide a high-resolution and
accurate thermal profile capable of capturing all hot spots are called Direct
Numerical Simulations (DNSs). These methods include the Finite Difference (FDM),
Finite Element (FEM), and Finite Volume Methods (FVM). The drawback of these
methods is that they require a very large degree of freedom for simulation and
therefore carry a high computational cost~\cite{jiang_2022_tcad}. This makes the
use of these methods challenging for chip-level thermal simulation.  Another
method of thermal simulation that is more efficient and more typically used, is
the RC-element thermal circuit. These models are commonly used for large-scale
CPUs due to their efficiency and simplicity~\cite{chaparro_2007,wu_2005}. RC
circuit models treat each Functional Block (FB) as a thermal node; thus, the
temperature in each FB is assumed to be uniform in space. This assumption limits
the potential accuracy of the model~\cite{jiang_2022}. This has been improved in
the HotSpot thermal simulator~\cite{hotspot_2003} by using a grid-based RC
circuit approach, but, even then, this method carries drawbacks in its accuracy
due to the inherent limitations of the thermal circuit approach.  Additionally,
the use of very small RC elements in grid mode causes the computational burden
of the RC circuit approach to become similar to DNS methods.

With the goal of achieving real-time TAS, the capabilities of the chosen thermal
model must be taken into consideration. A computationally heavy model, like a
DNS method, will limit the speed of the scheduling algorithm, leading to an
inability for the model to make decisions in real time. An inaccurate thermal
model could cause the algorithm to make sub-optimal decisions, or even to make
decisions that will damage the CPU while executing tasks.  Thus, there is a
requirement for a highly accurate and computationally efficient thermal model
that can be used in scheduling.

Proper Orthogonal Decomposition (POD), a physics-informed data-centric learning
model, can be applied in TAS. POD enables the complex thermal behavior of a
physical domain to be trained using a reduced-order method. This method uses
collected data to train POD modes that may be used for more efficient thermal
simulation of a CPU. Due to the reduced order inherent to POD, the computational
cost of thermal prediction is massively reduced. However, even with this
reduction in computational cost, the accuracy of the model is comparable to
DNS~\cite{jiang_2022}. In this work, we have performed an initial result
comparing POD prediction with FEM DNS and a dynamic RC circuit simulation
(HotSpot).  See Figure~\ref{fig:podhotfem} where POD outperforms the HotSpot in
its accuracy and very closely matches the FEM DNS.

Many current TAS algorithms, including RT-TAS~\cite{lee_2019}, are designed to utilize a steady
state thermal model~\cite{lee_2019,minimizing_2018,sha_2019}.  These models do
not provide information regarding the transient thermal state of the chip (See
Figure~\ref{fig:dynvsss}).  The transient thermal state of a chip depicts how
the temperature of the chip changes over time.  The transient thermal state can
vary from the steady state temperature of the chip due to the time it takes the
chip to reach the steady state temperature after a change in power dissipation.
Thus, in this work, we develop a dynamic POD-based TAS algorithm, named POD-TAS.
POD-TAS uses a reduced-order thermal modeling approach which enables a dynamic
model that is capable of accurate prediction of the transient thermal state with high
spatiotemporal resolution. This enables POD-TAS to assign tasks based on the
CPU's transient thermal behavior.

We implement the POD thermal model and POD-TAS scheduling algorithm and
replicate the steady-state RC circuit-based thermal model used by Lee et
al.~\cite{lee_2019} and the RT-TAS~\cite{lee_2019} scheduling algorithm for
comparative evaluation. This algorithm is chosen to compare to POD-TAS due
to the differences in the methods the algorithms employ for temperature
management. RT-TAS focuses on minimizing temperature by balancing the
steady-state temperatures of the CPU cores, while POD-TAS observes the transient
thermal state of the CPU. By comparing POD-TAS with an algorithm that relies
only on steady-state temperature, the potential advantages of using a dynamic
thermal model for scheduling can be investigated.

Using a simulation-based evaluation methodology (Section~\ref{sec:methods}), we
compare our proposed POD-TAS algorithm to RT-TAS~\cite{lee_2019}, an existing
state-of-the-art TAS algorithm that relies on a steady-state thermal model. Our
evaluation methodology enables a direct comparison of the CPU thermal behavior
while executing task schedules created by the two algorithms. Our simulation
pipeline includes gem5~\cite{gem5}, a cycle-accurate CPU simulator that allows
for detailed traces of architecture-level events during execution to be
collected.  These traces are then used as input to McPAT~\cite{mcpat_2009}, a
power simulator that can generate a dynamic power map of the power of each FB of
the CPU with high temporal resolution. Lastly, FEniCS~\cite{fenics_2003} is used
to collect the dynamic temperature profile of the CPU during schedule execution
via FEM-based DNS. Refer Figures~\ref{fig:fullflow},~\ref{fig:modelcreation},
and~\ref{fig:benchandsched}. The COMBS~\cite{combs_2020} benchmark suite is used
to provide task workloads during our evaluation of both of the TAS algorithms
(See Table~\ref{tab:benches}).

\textbf{Therefore, the contributions of this work can be summarized as follows:}
\begin{enumerate}
    \itemsep0em

    \item We implement a new TAS algorithm, POD-TAS, that
        controls the temperature of a multi-core CPU based on a pair of
        temperature thresholds (Section~\ref{sec:algo}).
        This algorithm is the first to use a POD-based
        thermal model, which provides temperature
        information with high spatiotemporal resolution.  We
        describe a set of CPU core thermal states and
        transitions among them to manage the CPU behavior,
        which avoids the assignment of tasks to cores in
        unsuitable states.  See Figure~\ref{fig:cpustates}.
        
    \item We compare the accuracy of a popular RC thermal circuit simulator, HotSpot, with
        a POD-based thermal model using FEM DNS as a baseline. See
        Figure~\ref{fig:podhotfem}. This comparison supports the use of a POD-based
        thermal model over others.

    \item This work provides a novel simulation-based methodology for the evaluation of
        TAS algorithms which measures the algorithm performance with high fidelity without
        the need for hardware setups (Section~\ref{sec:methods}).  A state of the art
        benchmark suite, COMBS~\cite{combs_2020}, is used to provide CPU workloads during
        task scheduling.

    \item We evaluate the POD-TAS algorithm to demonstrate its capability to manage the
        thermal behavior of the CPU. 

    \item POD-TAS is compared to the existing state of the art, RT-TAS~\cite{lee_2019},
        using the proposed methodology (Section~\ref{sec:methods}).  We implement the POD
        thermal model and POD-TAS scheduling algorithm and replicate the
        RT-TAS~\cite{lee_2019} thermal model and scheduling algorithm for comparative
        evaluation (See Figure~\ref{fig:results}).  Two evaluation cases per algorithm
        (POD-TAS and RT-TAS~\cite{lee_2019}) are performed using subsets of the benchmark
        suite, one with 4 benchmarks and another using 8 benchmarks. See
        Tables~\ref{tab:benches} and~\ref{tab:diffs}.
    
    \item Using a set of 4 benchmarks, we observe that POD-TAS can decrease the
        peak spatial thermal variance by 53.0\% and the peak chip temperature of 29.01\%.
        The variance of the maximum CPU temperature over time is decreased by 96.18\% and
        the variance of the mean temperature is decreased by 88.69\% as well.  Using a set of 8
        benchmarks, POD-TAS decreases the variance of the maximum chip temperature by
        93.26\%, and the peak chip temperature by 26.26\%. See Table~\ref{tab:diffs}.
    
\end{enumerate}

Our results indicate that the use of an accurate and
efficient thermal model, POD, can improve the performance of
TAS algorithms. However, the TAS algorithm must be designed
to leverage the capabilities of the dynamic thermal model
that underlies its decision-making.

The remainder of this paper is structured as follows:
Section~\ref{sec:related} reviews studies related to this
work and Section~\ref{sec:pod} describes the POD thermal
model used by POD-TAS.
Section~\ref{sec:models} investigates the
differences between POD and other common thermal models.
The effects that these differences may have on scheduling are
also discussed.
Section~\ref{sec:algo} provides a
formal description of the POD-TAS algorithm, while
Section~\ref{sec:methods} describes our evaluation
methodology.  Section~\ref{sec:eval} provides the results of
the evaluation of POD-TAS and its comparison with
RT-TAS~\cite{lee_2019}. Section~\ref{sec:future} identifies
potential future research directions to extend this study
and lastly, Section~\ref{sec:conc} concludes the paper.

\section{Related Work}
\label{sec:related}

The criteria of selecting existing state of the art that are
related to our study are to choose the most recent works
that have performed TAS to examine the methods by which they
manage the chip temperature and evaluate their TAS
algorithms.  The evaluation methodologies of TAS works tend
to use either a simulation
environment~\cite{minimizing_2018,shehzad_2019,tamcf_2019,rodriguez_2019,moulik_tarts_2021,rodriguez_2021,safari_2021,sharma_2022,akbar_2022,rodriguez_2022},
or a hardware set
up~\cite{lee_2019,kim_2020,maity_2021,benedikt_2021,bashir_2022,ozceylan_2022}.
While hardware can be argued to be very realistic;
simulation-based methodologies offer greater flexibility
without the risk of damaging the hardware during evaluation.
Another varying factor in TAS research is the thermal model
used for predicting/detecting the CPU temperature during
scheduling.  Some works rely on inbuilt temperature sensors
to log the real temperature~\cite{kim_2020,maity_2021},
while others~\cite{shehzad_2019} use the Analytical model of
Temperature in MIcroprocessors (ATMI) model, neural network
models~\cite{akbar_2022}, or a custom Ordinary Differential
Equation (ODE)-based methodology~\cite{ozceylan_2022}.
However, the most commonly used thermal model is the thermal
RC circuit.  Some works use the HotSpot simulator to
generate the RC circuit
model~\cite{moulik_tarts_2021,safari_2021,sharma_2022},
while others create their own RC
model~\cite{minimizing_2018,tamcf_2019,lee_2019,rodriguez_2019,benedikt_2021,rodriguez_2021,rodriguez_2022}.
The thermal models used in TAS algorithms are typically
either dynamic models that allow for the transient thermal
state to be predicted, or steady-state models that predict
the steady-state temperature of the CPU for a given power
dissipation. Additionally, some algorithms collect the real
temperature of the CPU using inbuilt temperature sensors.  A
lot of recent studies utilize DVFS to control the heat
output of the
CPU~\cite{minimizing_2018,kim_2020,maity_2021,moulik_tarts_2021,rodriguez_2021,bashir_2022,rodriguez_2022,sharma_2022},
or other
methods~\cite{shehzad_2019,tamcf_2019,lee_2019,rodriguez_2019,benedikt_2021,safari_2021,akbar_2022,ozceylan_2022}.
Table~\ref{tab:lit_rev} describes related state of the art
on TAS algorithms highlighting, for each study, the year of
publication, CPU model, thermal model, temperature type from
the thermal model, whether DVFS is used, evaluation
methodologies, and performance metrics. 

There are existing studies that aim to create TAS algorithms for mixed criticality systems, which
have low priority and high priority operating modes.  The algorithm by Li et
al.~\cite{minimizing_2018} uses an RC circuit model and can be tuned to focus on either energy aware
scheduling or thermal aware scheduling.  Li et al.~\cite{tamcf_2019} extend the existing algorithm
to support mixed-criticality systems by assigning each task a different Worst Case Execution Time
(WCET) for each criticality mode.  In the mixed criticality system-based algorithm, if a task cannot
complete execution while the system is in low criticality mode, then all low priority tasks are
dropped and the system switches to high criticality mode until its return to low criticality mode is
determined to be safe. The algorithm focuses on setting the execution rates for each task to
approximate a fluid schedule.  Safari et al.~\cite{safari_2021} also design a TAS algorithm for
mixed criticality systems, which relies on thermal safe power, defining the amount of power that a
processor can dissipate before it overheats.  Based on this defined value, it determines the maximum
safe number of simultaneously active cores that can be used for scheduling.

Akbar et al.~\cite{akbar_2022} use a neural network to predict the temperature of servers in a data
center to schedule tasks among a large set of machines.  The  TAS algorithm implemented by Sharma et
al.~\cite{sharma_2022} creates groups of CPU cores called clusters.  The algorithm initially assigns
tasks to the clusters, then the schedule is modified based on whether energy or temperature
efficiency is being prioritized.  Bashir et al.'s~\cite{bashir_2022} algorithm relies on DVFS and a
modified RC circuit model. Their algorithm controls the CPU power management states of the CPU to
regulate the static power dissipation of cores.

Many studies focus on heterogeneous processors, as they provide more task allocation options to a
TAS algorithm.  Benedikt et al.~\cite{benedikt_2021} design the MultiPAWS algorithm for an avionics
environment.  Their algorithm relies on a steady state RC circuit model to schedule tasks on an ARM
big.LITTLE CPU.  Kim et al.~\cite{kim_2020} also design a TAS algorithm for a heterogeneous ARM
big.LITTLE CPU which focuses on minimizing power dissipation by deciding whether to use DVFS or
migrate tasks between the big and LITTLE cores.  Maity et al.~\cite{maity_2021} implement their TAS
algorithm for an ARM big.LITTLE heterogeneous chip which includes a GPU.  Their algorithm uses the
LITTLE cores to operate the operating system, scheduler, and temperature sensors; on the other hand,
tasks are run on the big cores and GPU.  This is one of the works that uses inbuilt temperature
sensors during scheduling.  Lee et al. implement the Real-Time Thermal Aware
Scheduling (RT-TAS~\cite{lee_2019}) algorithm for a heterogeneous chip with an integrated GPU and CPU cores.  RT-TAS~\cite{lee_2019}
assigns tasks with the goal of maintaining a balanced steady state temperature among the cores. The
maintenance of this balance aims to minimize the chip temperature.  Our POD-TAS algorithm can be
readily expanded to support scheduling tasks on chips with an integrated GPU and CPU cores or other
heterogeneous chips.

There are studies that limit the execution rate of tasks using DVFS and other tools to control the
thermal behavior of the chip.  Ozceylan et al.~\cite{ozceylan_2022} create an algorithm that uses a
differential equation-based thermal model to predict the system temperature.  After finding the
minimum utilization required to complete a task by its deadline, the cpulimit tool is used
to limit the CPU speed.  Shehzad et al.~\cite{shehzad_2019} use an Earliest Deadline First (EDF)
methodology along with a fairness metric to create a TAS algorithm that aims to approximate a fluid
schedule by maintaining equal amounts of execution among tasks. These methods rely on limiting
the execution rate of tasks to control the CPU temperature and may face difficulty under a
heavy or sporadic task load. Additionally, not all chips support the use of DVFS for limiting
the execution rate of tasks.

Rodr\'iguez et al.~\cite{rodriguez_2019,rodriguez_2021,rodriguez_2022} create and expand algorithms
that rely on two temperature thresholds.  Their algorithm for single core processors aims to
maintain the CPU temperature below a maximum temperature threshold. This algorithm has two settings
that define whether it cools the CPU enough to execute the next task, or if after executing the task
it cools back down to the low threshold~\cite{rodriguez_2019}.  The existing algorithm is expanded
by incorporating DVFS to control the heating rate of the CPU~\cite{rodriguez_2021}.   Furthermore,
this study expands their previous works to support dual-core CPUs~\cite{rodriguez_2022}.  The
proposed POD-TAS algorithm differs from the algorithms developed by Rodr\'iguez et al. in their
conditions for inserting idle time for CPU cores and the prioritization of tasks and cores during
task assignment. Additionally, unlike Rodr\'iguez et al., the POD-TAS algorithm is not limited to
single and dual core CPUs and does not rely on DVFS for temperature control.

\begin{table*}[h]
    \centering
    \caption{Comparative literature review of selected recent studies}
    \label{tab:lit_rev}
    \scriptsize
        \begin{tabular}{|C{0.05\textwidth}|c|c|c|c|c|c|L{0.26\textwidth}|}
        \hline
            Ref.                &Year& \multirow{2}{*}{CPU Model}   & Thermal   & Temperature & Uses  & Evaluation & \multirow{2}{*}{Metrics Used} \\ 
        Num.                &    &             &  Model    & Type    & DVFS? & Method     &         \\ \hline
        \cite{shehzad_2019} &2019&  Custom     & ATMI      & Transient &       & Simulation & Transient Temp., Thermal Gradient, Max Spatial Gradient, Mean Core Temp.  \\ \hline
        \cite{tamcf_2019}   &2019&  Custom     & RC        & Transient &       & Simulation & Transient Temp., Energy, Acceptance Ratio  \\ \hline
        \cite{lee_2019}     &2019&Tegra X1     & RC        & Steady-State  &       & Hardware   & Peak Temp., Response Time, Temp. Vs. Utilization, Schedulability, Transient Temp. \\ \hline
        \cite{rodriguez_2019}&2019& Custom     & RC        & Transient &       & Simulation & Response Time, Schedulability \\ \hline
        \cite{kim_2020}     &2020& Exynos 5422 & Sensor    & Real    &  yes  & Hardware   & Normalized Execution Time, Mean Temp., Peak Temp., Thermal Emergencies        \\ \hline
        \cite{maity_2021}   &2021& Exynos 5422 & Sensor    & Real    &  yes  & Hardware   & Peak Temp.        \\ \hline
        \cite{benedikt_2021}&2021& NXP i.MX8   & RC        & Steady-State  &       & Hardware   & Steady State Temp., Schedule Visualizations     \\ \hline
        \cite{rodriguez_2021}&2021& Custom     & RC        & Transient &  yes  & Simulation & Response Time, Min/Mean/Max Temp. \\ \hline
        \cite{safari_2021}  &2021&ARM Cortex-A15&RC(HotSpot)&Steady-State &       & Simulation & Feasibility, Schedulability, Thermal Violations, Quality of Service, Peak Temp., Reliability \\ \hline
        \cite{sharma_2022}  &2022& Custom ARMv8& RC(HotSpot)&Steady-State  &  yes  & Simulation & Mean Temp. Success Ratio, Energy Consumption, Context Switching \\ \hline
        \cite{akbar_2022}   &2022& Data Center &  DNN      & Steady-State &       & Simulation & Energy, Temp. Difference, Migrations, Service Level Agreement Violations \\ \hline
        \cite{bashir_2022}  &2022& Xeon 2680v3 &RC(modified)&Transient &  yes  & Hardware   & Power, Mean Temp., Schedulability, Overheats        \\ \hline
        \cite{rodriguez_2022}&2022& Custom     & RC        & Transient &  yes  & Simulation & Response Time, Min/Mean/Max Temp. \\ \hline
        \cite{ozceylan_2022}&2022& Exynos 5422 & ODE-based & Transient &       & Hardware   & Max Temp., Variance  \\ \hline
            \hline
        \textbf{This work}  & -- &Athlon II X4 640& POD    & Transient &       & Simulation & Peak Temp., Peak Variance, Var(Mean), Var(Max), Var(Var)        \\ \hline
    \end{tabular}
\end{table*}

Table~\ref{tab:lit_rev} highlights the summarized
    information about the reviewed studies in comparison to
    ours.  As we can see from Table~\ref{tab:lit_rev}, many
    studies on thermal-aware real-time scheduling rely on RC
    circuit-based thermal models, which carry an isothermal
    assumption for each RC element.  This assumption can
    lead to the calculation of incorrect heat flux across
    the element interfaces, which is worse in the dynamic
    case.  Due to the assumption of isothermic elements in
    the RC circuit, a popular RC-based thermal simulator,
    HotSpot~\cite{hotspot_2003}, is observed to have an
    error as large as 200\% in its block mode, when compared
    to FEM analysis~\cite{jiang_2022}. This lack of accuracy
    can cause a TAS algorithm to make sub-optimal decisions
    during scheduling that may result in hardware damage due
    to overheating. 

Furthermore, as we can see from Table~\ref{tab:lit_rev}, 6
out of the 14 reviewed TAS methods rely on DVFS.  While
DVFS can provide more control over the chip behavior, it
limits the generalizability of an algorithm, as the
algorithm will only be usable on processors that support
DVFS.  Algorithms that do not rely on DVFS are able to
perform scheduling for chips that do not support DVFS,
as well as those that do by maintaining a constant
frequency. Thus, POD-TAS is designed to maintain generalizability by
avoiding DVFS. 

\section{Physics-Informed Thermal Modeling Enabled By Proper Orthogonal Decomposition~\cite{jiang_2022_iscas}}
\label{sec:pod}

POD is a physics-informed model which is able to represent a
complex thermal problem in both time and space using trained
POD modes, together with the Galerkin Projection (GP) of the
heat equation~\cite{jia2015fast,jia2014thermal}.  The modes
are extracted from thermal solution data of the physical
domain in a training process. The data used for training
should include a range of parametric variations to inform
the model about the response of the physical quantity in the
domain in a range of conditions.  After training, the modes
created are tuned to the parametric variations, such as
Boundary Conditions (BCs) and power variations. The GP futher
provides a physics-based guidance to improve the model accuracy and
efficiency.

\subsection{Construction of POD Modes}

The POD modes are optimized by maximizing the mean square
inner product of the thermal solution data with the modes
over the entire
domain~\cite{lumley1967structure,berkooz1993proper} while
taking into account the dynamic or static parametric
variations of BCs and interior power sources. In our study,
the thermal data is collected at each simulation time step
in the FEM DNS (Section~\ref{sec:intro}) for the data collection.  This optimization
process~\cite{lumley1967structure,berkooz1993proper} leads
to an eigenvalue problem, described by the Fredholm
equation, shown in Eq.~\eqref{eq:2}:
\begin{equation}
   \int_{{ \Omega}}R(\vec{r},\vec{r}')  \varphi (  \vec{r} ' ) d  \vec{r} '  =  \lambda \varphi( \vec{r } ),  \label{eq:2}
\end{equation}
where $\lambda$ is the eigenvalue corresponding to the POD
mode (i.e., eigenfunction $\varphi$)  and $R(\vec{r},
\vec{r}')$ is a two-point correlation tensor. This two-point
correlation tensor is shown in Eq.~\eqref{eq:3}:
\begin{equation}
R( \vec{r},   \vec{r} '  )= \langle T(  \vec{r} , t)  \otimes T(  \vec{r}' , t)\rangle\label{eq:3}
\end{equation}
with $\otimes$ as the tensor operator and $\langle \rangle$
indicating the average over the number of thermal data sets.
With the modes obtained from the training process, including
collection of the thermal data and solving the eigenvalue
problem as shown in Eq.~\eqref{eq:2}, the temperature can then be
represented by Eq.~\eqref{eq:4}:
\begin{equation}
    T( \vec{r}, t ) =  \sum_{i=1}^M  a_{i}(t)  \varphi _{i}( \vec{r} ),\label{eq:4}
\end{equation}
where $M$ is the number of POD modes chosen to reconstruct
the temperature solution.

\subsection{Projection of the Thermal Problem to POD Space}

The GP is used to construct a POD model by
projecting the heat transfer equation onto POD space, as shown in Eq.~\eqref{eq:5}:
\begin{equation}
    \int_ \Omega (\varphi _{i}( \vec{r} ) \frac{\partial  \rho CT}{\partial t} +  \nabla    \varphi _{i}(\vec{r})   \cdot  k  \nabla  T ) d \Omega   =  \int_ \Omega    \varphi _{i} ( \vec{r} )  P_{d} ( \vec{r}, t ) d \Omega -  \int_S \varphi _{i} ( \vec{r} )(-k  \nabla  T \cdot   \vec{n})dS,
    \label{eq:5}
\end{equation}
where $k$, $\rho$, and $C$ are the thermal conductivity,
density, and specific heat, respectively. $P_{d} ( \vec{r},
t)$ is the interior power density, $S$ is the boundary
surface, and $\vec{n}$ is the outward normal vector of the
boundary surface. With the selected POD modes, Eq.~\eqref{eq:5}
can be rewritten as an $M$-dimensional Ordinary Differential
Equation (ODE) for $a_{i}(t)$ as shown in Eq.~\eqref{eq:6}:
\begin{equation}
     \sum_{i=1}^M  c_{i,j} \frac{da_{i}(t)}{dt} +\sum_{i=1}^M  g_{i,j} a_{i}(t) =  P_{j}, \: j = 1\: \textrm{to}\:M, \label{eq:6}
\end{equation}
where $c_{i,j}$ and $g_{i,j}$ are the elements of thermal
capacitance matrices and thermal conductance matrices in the
POD space and they are defined in Eq.~\eqref{eq:7} and Eq.~\eqref{eq:g} respectively as:
\begin{equation}
    c_{i,j}=\int_\Omega\rho C\varphi_{i}(\vec{r})\varphi_{j}(\vec{r})d\Omega \label{eq:7}
\end{equation}
\begin{equation}
    \text{and}\:\:g_{i,j}=\int_\Omega k\nabla\varphi_{i}(\vec{r})\cdot\nabla\varphi_{j}(\vec{r})d\Omega \label{eq:g}
\end{equation}
$ P_{j}$ in Eq.~\eqref{eq:6} is the power source strength for
the $j$-th POD mode in the POD space and is described below in Eq.~\eqref{eq:8}:
\begin{equation}
    P_{j}= \int_ \Omega  \varphi _{j}(\vec{r}) P_{d} ( \vec{r}, t ) d \Omega -  \int_S \varphi _{j} ( \vec{r} )(-k  \nabla  T \cdot   \vec{n})dS.\label{eq:8}
\end{equation}
Once the power consumption is obtained, the interior power
source strength in POD space given in Eq.~\eqref{eq:8} can be
pre-evaluated. For the boundary heat source in Eq.~\eqref{eq:8},
the BC of the substrate bottom is modeled by convection heat
transfer with a constant heat transfer coefficient and an
ambient temperature of 45 $^\circ$C ($T_{amb}$). All other
boundaries are adiabatic. The coefficients in Eq.~\eqref{eq:7}
can also be pre-evaluated once the modes are determined.
With $a_{i}(t)$ solved from Eq.~\eqref{eq:6} in POD simulation,
the temperature solution can be predicted from Eq.~\eqref{eq:4}.

\section{Thermal Model Selection and Effects on Scheduling}
\label{sec:models}

\begin{figure}[b]
    \centering
    \includegraphics[width=0.5\textwidth]{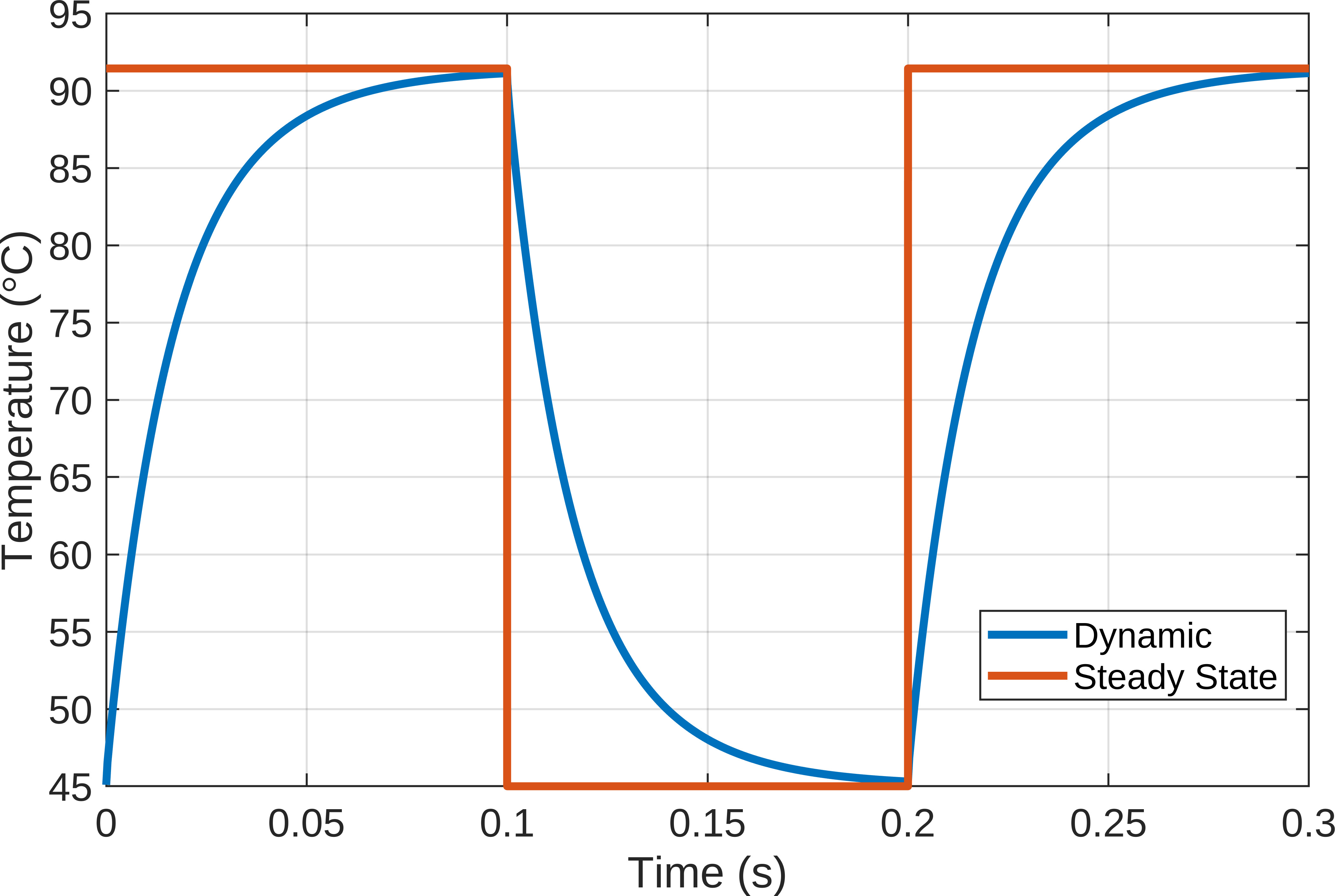}
    \caption{Dynamic temperature prediction compared to steady state}
    \label{fig:dynvsss}
\end{figure}

\begin{figure}
    \centering
    \includegraphics[width=0.97\textwidth]{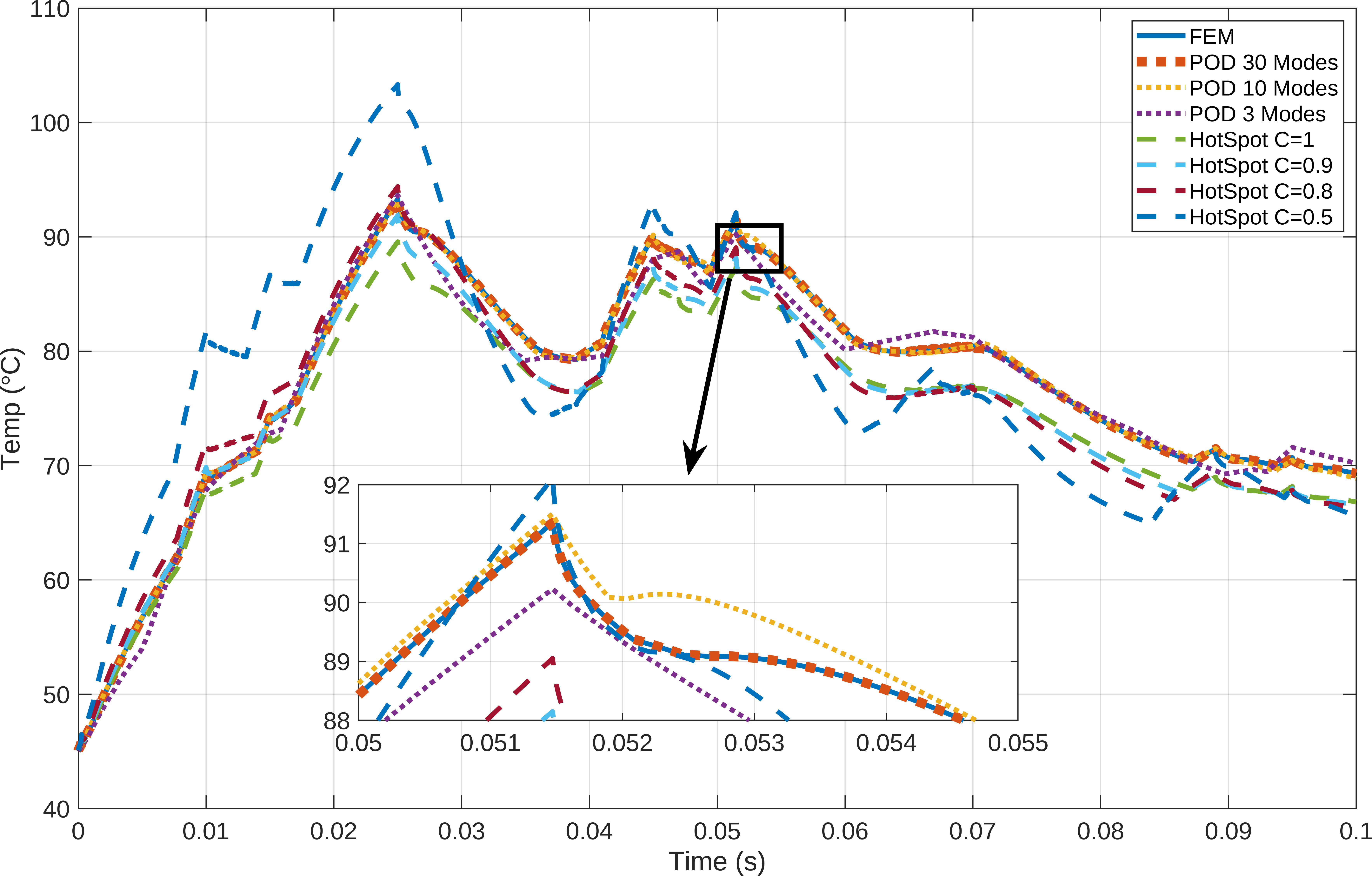}
    \caption{Comparing POD with HotSpot and FEM DNS}
    \label{fig:podhotfem}
\end{figure}

The effects that the chosen thermal model can have on the
performance of a TAS algorithm are manifold. An inaccurate
temperature prediction can lead to sub-optimal or
destructive decisions.  On the other hand, some thermal
models may not provide detailed enough information to make
optimal scheduling decisions. Thus, the choice of the thermal
model used when designing a TAS algorithm is a defining
factor for its overall performance.
In this section, we compare the differences between steady-state
and dynamic thermal models (Figure~\ref{fig:dynvsss}). Also, we
present an initial result (Figure~\ref{fig:podhotfem})
that compares our thermal model, POD, with HotSpot RC
thermal circuit simulation, and FEM DNS.

A dynamic thermal model predicts the temperature of the CPU
over time. On the other hand, a steady state model only
predicts the temperature at which the CPU will reach
equilibrium given a certain power dissipation.
Figure~\ref{fig:dynvsss} shows the predictions of a dynamic
thermal model alongside the predictions of a steady state
thermal model.  When the power dissipated is high, as shown
at $t=0$s and $t=0.2$s, the steady state model predicts a
temperature of 91.43$^{\circ}$C.  However, the core takes a
significant amount of time to approach that temperature, as
shown by the dynamic predictions. Given this difference, the
scheduler may make incorrect decisions based on the
steady-state predictions unless the decision rate of the
scheduler is extremely slow. However, as shown in
Table~\ref{tab:benches}, the execution time for most tasks
is far less than the time needed for the CPU to approach
steady-state temperature. 
In this case, extra slack time may be generated for a task that does
not need it. Due to the lack of transient thermal
prediction, steady-state thermal models limit the ability of
an algorithm to have a precise knowledge of the current
thermal state of the CPU it is currently scheduling tasks
for. This limits the ability of the algorithm to be able to
make optimal decisions regarding task scheduling.
Thus, dynamic temperature
prediction is required for a TAS algorithm to make
well-informed decisions based on the CPU thermal state at a
given time.

A popular RC thermal circuit simulator is
HotSpot~\cite{hotspot_2003}.  This simulator allows for
simulation in two modes, the first of which is block mode,
where each functional unit of the CPU is treated as an RC
node. The second mode is grid mode, which allows for the
spatial domain to be treated as a mesh of RC elements,
allowing for much higher simulation accuracy at the cost of
computation time.  This simulator also includes a scaling
factor setting which scales the capacitance of lumped
thermal nodes to aid in simulation accuracy. 

We compare POD with FEM DNS, a rigorous and
accurate thermal simulation~\cite{jiang_2022_tcad}, and with
HotSpot, a dynamic RC thermal circuit
simulator~\cite{hotspot_2003}. This comparison sheds light
on the accuracy of HotSpot and POD when compared with FEM
DNS as a baseline. To this end, FEM is configured with a
mesh size of 200$\times$200$\times$17 to allow for high
spatial resolution in the $x$, $y$, and $z$ directions. POD
is trained with, and thus predicts, temperature with the
same resolution and mesh size as FEM. HotSpot is limited to
only support $x$ and $y$ mesh sizes that are powers of 2, so
to support fair comparison between the models, we configure
it to use a mesh of 256$\times$256$\times$17. The spatial
domain in all cases is representative of an AMD Athlon II X4
640 CPU, which has a size of 14mm$\times$12mm$\times$0.3mm.

Figure~\ref{fig:podhotfem} is a plot of the temperature over
time where we use three different methods, POD, HotSpot, and
FEM DNS to simulate the CPU temperature.  All methods use
the same power input data for consistency. The FEM DNS is
the solid blue line in the plot, which is the baseline to
compare the other models to. The four dashed lines are the
temperature calculated using HotSpot with different scaling
factors, denoted as $C$, which is set to 1.0, 0.9, 0.8, and
0.5.  Lastly, the three dotted lines are predictions using
our POD model with varying numbers of modes: 3, 10, and 30.
We can see from the figure that HotSpot does not follow the
FEM temperature well and over time it diverges from the FEM
curve, unlike POD. From the inset plot in
Figure~\ref{fig:podhotfem}, we can see that as the number of
POD modes increases, the prediction accuracy greatly
increases. Around 0.052 seconds, the POD prediction with 30
modes is almost exactly following the FEM result, while the
10 modes result has some error, while the 3 modes result is
close to the FEM result, but we can see that the slope
differs greatly from the FEM result in this region.
However, even with only 3 modes, the POD prediction is more
accurate than the HotSpot result.  Using 30 POD modes, the
average prediction error of the maximum chip temperature is
0.0033\%, while with hotspot, the lowest error is 2.12\% at
C=0.8.  Therefore, we do not evaluate our proposed TAS
algorithm using a dynamic RC thermal circuit model such as
HotSpot due to its accuracy. Even though the temperature
output is very similar between POD with 30 modes and FEM
DNS, we do not perform TAS using FEM DNS due to its
computational cost~\cite{jiang_2022_tcad}.

Furthermore, as shown in Figure~\ref{fig:podhotfem},
different thermal models and differing configurations can
provide very different accuracy. For instance, we can see
that HotSpot, when configured with a scaling factor of 0.5
predicts the temperature to be much higher than the FEM
baseline at 0.02 seconds. In this case, the scheduling
algorithm that relies on HotSpot would generate extra slack
time for that CPU core even though in reality, it is not
needed. Conversely, at 0.035 seconds, we can see that many
of the HotSpot temperature curves are below the ground-truth
curve. In this case, the algorithm would receive the
incorrect information that the CPU has more thermal capacity
available than it realistically does, leading to over
heating of the device. Thus, a highly accurate thermal model
is required to guarantee the safety of the hardware when
using a TAS algorithm.

\section{A TAS Algorithm Based on a Dynamic POD Thermal Model}
\label{sec:algo}

As shown, different thermal models can provide temperature
prediction with a wide variety of characteristics. Given the
high performance of the POD-based thermal model, this model
is chosen as the basis for our POD-TAS algorithm. Thus, the
algorithm should be designed to leverage the dynamic, high
resolution thermal prediction of the model.
The POD-TAS algorithm leverages a pair of temperature thresholds to restrain the maximum
core temperature. Once tasks are selected and assigned, POD (Section~\ref{sec:pod}) is
used to predict the temperature of all cores of the CPU so that the cores may be idled
when the temperature within it exceeds the threshold. Once the predicted temperatures are
observed to reach the temperature threshold, cores exceeding the hot temperature
threshold, $T_H$, are idled. The idling continues until the core reaches a ``cool''
threshold, $T_C$, and is then allowed to resume execution.

\begin{table}
    \centering
    \caption{Table of Notations}
    \label{tab:notat}
    \begin{tabular}{|c|c|}
        \hline
        \textbf{Notation} & \textbf{Meaning} \\ \hline
        $\Delta t$      & The time step size used during scheduling \\ \hline
        $\tau$          & Set of Tasks \\ \hline
        $\pi$           & Set of CPU cores \\ \hline
        $\Lambda$       & Mapping of tasks to cores \\ \hline
        $\Psi$          & Set of CPU states \\ \hline
        $\Psi_{\pi_i}$  & The state of the CPU core $\pi_i$ \\ \hline
        $T_H$           & The high temperature threshold \\ \hline
        $T_C$           & The low temperature threshold \\ \hline
        $T_{\pi}$       & The temperature of the CPU $\pi$ \\ \hline
        $T_{\pi_i}$     & The temperature of the CPU core $\pi_i$ \\ \hline
        $n()$           & A function that determines the number of elements in a set \\ \hline
    \end{tabular}
\end{table}

\begin{algorithm}[t]
    \caption{Pseudocode of POD-TAS Algorithm (Primary Concept)}
    \label{alg:podtas}
    \begin{algorithmic}[1]
        \Function{POD-TAS}{$\tau$, $T_C$, $T_H$, $t_{end}$}
            \State $t_{curr}\gets0$;
            $\sigma\gets[]$;
            \While{$t_{curr}<t_{end}$}
                \State $\tau'\gets\text{SELECT\_TASKS}(\tau,\pi,\Psi)$
                \State $\Lambda\gets\text{ASSIGN\_TASKS}(\tau', \pi, \Psi)$
                \State $\text{\textit{append}}(\sigma, (t_{curr},\Lambda))$
                \Do
                    \State $t_{curr}\gets t_{curr}+\Delta t$
                    \State $T_{\pi}\gets\text{PREDICT\_TEMP}(P_{\sigma},t_{curr})$
                    \State $\Psi\gets\text{UPDATE\_STATES}(\pi,\Psi,T_{\pi},\Lambda)$
                \doWhile{$\text{max}(T_{\pi_i}(t_{curr}),\forall\pi_i\in\pi)<T_H\text{ \textbf{and} }t_{curr}<t_{end}$}
            \EndWhile
            \State \Return $\sigma$
        \EndFunction
    \end{algorithmic}
\end{algorithm}

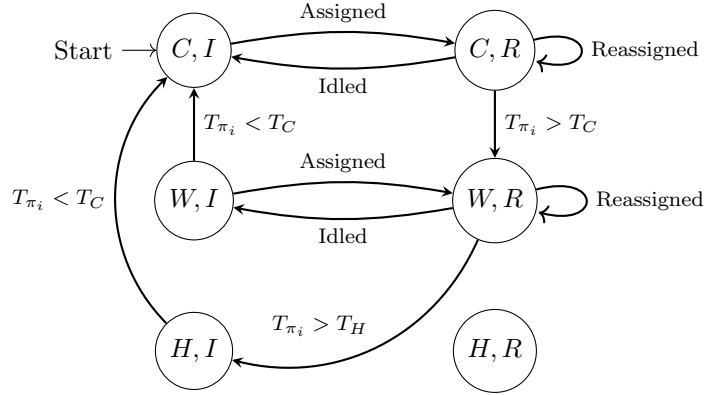
\begin{figure}[ht]
    \centering
    \begin{tikzpicture} [node distance = 2cm, on grid, auto]
        \node (ci) [state, initial, initial text = {Start}] {$C,I$};
        \node (cr) [state, right =4cm of ci] {$C,R$};
        \node (wi) [state, below = of ci] {$W,I$};
        \node (wr) [state, below = of cr] {$W,R$};
        \node (hi) [state, below = of wi] {$H,I$};
        \node (hr) [state, below = of wr] {$H,R$};

    \path [-stealth, thick]
        (ci) edge [bend left=10] node {\footnotesize Assigned} (cr)
        (cr) edge [bend left=10] node {\footnotesize Idled} (ci)
        (cr) edge [loop right] node {\footnotesize Reassigned} ()
        (wi) edge [bend left=10] node {\footnotesize Assigned} (wr)
        (wr) edge [bend left=10] node {\footnotesize Idled} (wi)
        (wr) edge [loop right] node {\footnotesize Reassigned} ()
        (cr) edge node {\footnotesize $T_{\pi_i}>T_C$} (wr)
        (wr) edge [bend left=40] node [above left] {\footnotesize $T_{\pi_i}>T_H$} (hi)
        (wi) edge node [right] {\footnotesize $T_{\pi_i}<T_C$} (ci)
        (hi) edge [bend left=45] node {\footnotesize $T_{\pi_i}<T_C$} (ci)
        ;
    \end{tikzpicture}

    \caption{State Transition Diagram for CPU Cores in POD-TAS}
    \label{fig:cpustates}
\end{figure}

Algorithms~\ref{alg:podtas},~\ref{alg:states},~\ref{alg:select},
and~\ref{alg:assign} represent the primary concept of our
POD-TAS algorithm. Based on these concepts, we have
implemented our novel POD-TAS algorithm into a scheduling
system which enables use of the algorithm to generate
schedules in a computer-readable format for use in
evaluation.
A description of the use of the scheduling
system is provided in Section~\ref{sec:methods}.  In our
POD-TAS algorithm, the notation $\pi$ represents a CPU
platform, with $\pi=\{\pi_1, \pi_2, ... , \pi_m\}$ being the
set of $m$ CPU cores. The temperatures at time $t$ are
$T_{\pi_i}(t),\pi_i\in\pi\text{ where }1\leq i\leq m$. The
set of tasks is denoted as $\tau$, such that $\tau_k\in\tau$
represents the $k$-th task. The mapping or assignment of the
tasks from $\tau$ to the CPU cores in $\pi$ is represented
as $\Lambda$.  $\Lambda_{\pi_i}$ represents the task that is
assigned to CPU core $i$ where $\pi_i$ belongs to a sorted
subset of $\pi$ that have valid thermal states. The sorting
of the elements of $\pi$ is explained later in this section.
The notation $\Psi$ in our algorithm represents the set of
CPU core states as described in Figure~\ref{fig:cpustates}.
We use $\Psi_{\pi_i}$ to represent the state of the $i$-th
CPU core, $\pi_i$.  The primary concept of the assignment of
cores depending on their thermal state is represented in
Algorithm~\ref{alg:states}. The notation $\sigma$ is used to
describe an ordered set of task-to-core assignments paired
with their assignment times (schedule).  Additionally, to
denote the power dissipated by a certain schedule, we use
the notation $P_{\sigma}$. The power being dissipated by the
CPU cores is dependent on the task assignment, and the power
dissipated by each task at a given time instant is dependent
on how much run time has been given to the task by the
schedule $\sigma$. Thus, $P_{\sigma}$ contains the power
dissipation of the CPU while it is running schedule $\sigma$
over time.  To make the powermap $P_j$ in Eq.~\ref{eq:6}
(Section~\ref{sec:pod}), the algorithm constructs
$P_{\sigma}$ to be used in each POD prediction.

To denote the thermal state of the CPU cores, we describe a
set of states and transitions among them to manage the CPU
behavior.  Figure~\ref{fig:cpustates} describes the possible
CPU states, $\Psi$, and their transitions. Each state
consists of a pair of letters of which the first letter
denotes the thermal state of the core and the second letter
denotes the execution state of the core. The possible
thermal states are $C$, $W$, and $H$, which correspond to
cool, warm, and hot respectively.  The possible execution
states are $I$ and $R$, where $I$ denotes an idle core, and
$R$ denotes a running core. The definition of the states of
each CPU core and the transitions between them defines how
the algorithm should treat CPU cores depending on their
state and what actions should be taken depending on their
state. In this way, we can avoid the assignment of tasks to
cores that are in unsuitable states. To our knowledge, no
other study has used such concept of core assignment
depending on its state.

\begin{algorithm}[h]
    \caption{Pseudocode of POD-TAS Core State Management Algorithm (Primary Concept)}
    \label{alg:states}
    \begin{algorithmic}[1]
        \Function{Update\_States}{$\pi$,$\Psi$,$T_{\pi}$,$\Lambda$}
            \For{$\pi_i\in\pi$}
                \State $run\_state\gets R$
                \If{$\Lambda_{\pi_i} = \emptyset$}
                    \State $run\_state\gets I$
                \EndIf
                \If{$T_{\pi_i}<T_C$}
                    \State $\Psi_{\pi_i}\gets(C,run\_state)$
                \ElsIf{$T_C<T_{\pi_i}<T_H$}
                    \State $\Psi_{\pi_i}\gets(W,run\_state)$
                \Else
                    \State $\Psi_{\pi_i}\gets(H,run\_state)$
                \EndIf
            \EndFor
            \State \Return $\{\Psi_{\pi_i}\:\:\text{where}\:\:\pi_i\in\pi\}$
        \EndFunction
    \end{algorithmic}
\end{algorithm}

Algorithm~\ref{alg:states} describes the subroutine
UPDATE\_STATES which includes the procedure of core state
management and is called in Algorithm~\ref{alg:podtas}.  As
shown in Figure~\ref{fig:cpustates}, all cores start in the
cool, idle state, $(C,I)$. From this state, when a task is
assigned to the core, it moves to the cool-running state,
$(C,R)$.  From the $(C,R)$ state, a core can return to
$(C,I)$ if the core is idled due to task completion or
reassignment. However, if the temperature of the core,
$T_{\pi_i}$ exceeds $T_C$, the core moves to the state
$(W,R)$, which denotes a warm, running core.  The warm
states denote that the core has crossed $T_C$ but has not
yet crossed $T_H$, thus the core is in a transitional state.
Similar to the transitions from $(C,R)$, a core in the
$(W,R)$ state can be idled, moving it to the $(W,I)$ (warm,
idle) state, but should its temperature exceed $T_H$, 
meaning a temperature violation has occured, the
core is forced to idle, and moved to the $(H,I)$ (hot, idle)
state.  Cores in the $(H,I)$ state are disallowed from
executing any tasks until their temperature falls below
$T_C$ and they return to state $(C,I)$.  This restriction
ensures that after a task exceeds the high temperature
threshold, it must cool down before resuming execution. This
ensures that no cores will rapidly alternate between task
execution and idling.  The $(H,R)$ (hot, running) state is
separate from the other states with no transitions to it
because that state is forbidden. As soon as the temperature
of a core is observed to exceed $T_H$, the core is idled.

Algorithm~\ref{alg:podtas} formally describes the primary
concept of the POD-TAS algorithm. The algorithm takes as
input $\tau$, the set of tasks to be ran; $T_C$, the CPU
cool temperature threshold; $T_H$, the CPU hot temperature
threshold; and $t_{end}$, the amount of time that the
schedule should be constructed for. The function returns the
schedule, $\sigma$, which can be stored for use by a
scheduling system. There are four subroutines that are
called: UPDATE\_STATES, SELECT\_TASKS, ASSIGN\_TASKS, and
PREDICT\_TEMP.  PREDICT\_TEMP takes the dynamic power map of
the current schedule, and the time for which to predict, and
predicts the CPU temperature for each CPU core at that time
using the POD thermal model. UPDATE\_STATES is described in
Algorithm~\ref{alg:states}, which is explained before with
the help of Figure~\ref{fig:cpustates}.  SELECT\_TASKS and
ASSIGN\_TASKS are described in the remainder of this
section.

\begin{algorithm}[h]
    \caption{Pseudocode of POD-TAS Task Selection Algorithm (Primary Concept)}
    \label{alg:select}
    \begin{algorithmic}[1]
        \Function{Select\_Tasks}{$\tau$,$\pi$,$\Psi$}
            \State $n\_tasks \gets n(\{ \pi_i \in \pi | \Psi_{\pi_i}\neq \Psi(H,I)\})$
            \State $\tau' \gets sort(\tau | TimeLeft(\tau_i) \geq TimeLeft(\tau_{i+1}))$
            \State $\tau_{run} \gets \{\}$
            \While{$n(\tau_{run}) < n\_tasks$}
                \State $\tau_{run} \gets \tau_{run} \cup \{\tau'[n(\tau_{run})]\}$
            \EndWhile
            \State \Return $\tau_{run}$
        \EndFunction
    \end{algorithmic}
\end{algorithm}

In POD-TAS, temperature management is handled primarily by
the two thresholds, but task selection from the queue and
core assignments can have a significant impact on the
schedule created and on the schedulability of a task set.
Task selection is handled by prioritizing the execution of
tasks with more remaining run time ($TimeLeft(\tau_i)$).
This helps to avoid tasks being idled unnecessarily and not
being able to be completed by their deadline. The task
selection algorithm (the subroutine SELECT\_TASKS) is
described in Algorithm~\ref{alg:select}. The primary concept
of this algorithm is that for each CPU core that is not in
the $(H,I)$ state, a task should be selected from $\tau$,
prioritizing tasks with more remaining execution time.

\begin{algorithm}[h]
    \caption{Pseudocode of POD-TAS Task to Core Assignment Algorithm (Primary Concept)}
    \label{alg:assign}
    \begin{algorithmic}[1]
        \Function{Assign\_Tasks}{$\tau_{run}$,$\pi$,$\Psi$}
            \State $\pi' \gets sort(\{\pi_i\in\pi|\Psi_{\pi_i}\neq \Psi(H,I)\} | T_{\pi_i} \leq T_{\pi_{i+1}})$
            \For{$\tau_k\in\tau_{run}$}
                \For{$\pi_i\in\pi'$}
                    \If{$\Lambda_{\pi_i}=\emptyset$}
                        \State $\Lambda_{\pi_i}\gets\tau_k$
                        \State $\text{break}$
                    \EndIf
                \EndFor
            \EndFor
            \State \Return $\{\Lambda_{\pi_i} \text{ where } \pi_i\in\pi'\}$
        \EndFunction
    \end{algorithmic}
\end{algorithm}

Algorithm~\ref{alg:assign} (the subroutine ASSIGN\_TASKS)
describes the assignment of selected tasks to available
cores.  This algorithm first sorts the cores in ascending
order of temperature so that cooler cores are assigned the
tasks requiring more execution time. This prioritization
scheme aims to provide the maximum amount of execution time
for the tasks that require it, since
Algorithm~\ref{alg:select} ensures that the tasks with the
highest remaining run time are prioritized. The interplay
and effects of these two algorithms is a potential future
research direction. This subroutine creates $\Lambda$,
which notes which tasks from the task set, $\tau$, should be running on 
each CPU core ($\pi_i$). If a CPU core, $\pi_i\in\pi$, does not
have an assignment, then $\Lambda_{\pi_i}$ will not exist for
that assignment, meaning that core is idle. This can occur
when either more cores are available in $\pi'$ than there are
tasks in $\tau_{run}$, or if $\pi_i$ is overheated, that is
$\Psi_{\pi_i}=(H,I)$.

We utilize Python as the main language for the POD-TAS implementation, with the
Numpy~\cite{numpy} package used to support mathematical operations. The ODE in
Eq.~\eqref{eq:6} must be solved to obtain $a_i(t)$ for POD temperature prediction.  The
PETSc~\cite{petsc} library is used to implement a program in C$++$ to solve the ODE. The
values of $c_{i,j}$ and $g_{i,j}$ for Eq.~\eqref{eq:6} are pre-calculated using
Eq.~\eqref{eq:7} and Eq.~\eqref{eq:g} respectively. To provide $P_j$, the dynamic power
map for each benchmark is arranged according to the schedule that has been constructed.
After constructing $P_j$, the main script writes out an output, which is loaded by the ODE
solver program to solve Eq.~\eqref{eq:6}.  The solution, $a_{i}(t)$, is then loaded
using Numpy~\cite{numpy} to solve Eq.~\eqref{eq:4} to obtain the predicted temperatures to
be used in the POD-TAS algorithm.

\subsection{Runtime Complexity of POD-TAS}

Algorithm~\ref{alg:podtas} is a loop with linear time
complexity based on $t_{end}/\Delta t$, which calls
Algorithms~\ref{alg:states},~\ref{alg:select},
and~\ref{alg:assign}. Algorithm~\ref{alg:states} uses a
linear loop through the CPU cores to update the core states
in $\Psi$. Algorithm~\ref{alg:select} sorts the tasks in
non-increasing order based on their remaining execution
times before using a linear time loop to take enough tasks
to fill the available CPU cores. The largest cost in
Algorithm~\ref{alg:select} is the sorting of the tasks based
on their remaining time, which using common efficient
sorting algorithms can be achieved in $\Theta(n\:log(n))$
time, where $n$ is the number of tasks to be sorted.
Algorithm~\ref{alg:assign} sorts the CPU cores based on
their temperatures before a pair of nested loops which
iterate through the selected tasks and the available CPU
cores. Thus, Algorithm~\ref{alg:assign} runs with
$\Theta(m\:log(m))+qr$ complexity, where $m$ is the number
of CPU cores in $\pi$, $q$ is the number of tasks in
$\tau_{run}$, and $r$ is the number of CPU cores in $\pi'$.
The number CPU cores can be assumed to be small in the
general CPU scheduling case, while the number of tasks does
not tend to be large enough to have a heavy impact on the
computational cost of sorting the set of tasks. Thus, the
primary source of computation is Algorithm~\ref{alg:podtas},
with its loop that calls the other three algorithms. So we
can say that POD-TAS overall has a linear time complexity
dependent on $t_{end}/\Delta t$. The RT-TAS~\cite{lee_2019} algorithm, which
we replicate and compare POD-TAS to, also has linear time
complexity~\cite{lee_2019}.

\subsection{Differences Between POD-TAS and RT-TAS~\cite{lee_2019} Algorithms}

The RT-TAS~\cite{lee_2019} algorithm supports the co-scheduling of an integrated CPU-GPU platform, such as
in a heterogeneous chip. The algorithm also supports platforms with only CPU cores, and no
integrated GPU. The evaluation performed by Lee et al.~\cite{lee_2019} demonstrates the
performance of the RT-TAS~\cite{lee_2019} algorithm on a computing platform with an integrated GPU over a
large period of time ($>17$ minutes). Thus, there is a need to replicate their algorithm
on a CPU platform without an integrated GPU to examine its performance in this
environment. Furthermore, the temperature sensors provided in a computing chip, which are
used in the RT-TAS~\cite{lee_2019} evaluation, are not capable of measuring the chip temperature with high
spatiotemporal resolution. Thus, a more detailed evaluation methodology is required. The
methodology used in this work (Section~\ref{sec:methods}) allows evaluation with
high spatiotemporal resolution.
Hence, we implement the POD thermal model and POD-TAS scheduling algorithm and replicate
the RT-TAS~\cite{lee_2019} thermal model and scheduling algorithm 
for comparative evaluation. Our evaluation includes a set of 4 tasks to provide a case
similar to the one used by RT-TAS~\cite{lee_2019}. Additionally, we evaluate with 8 tasks to examine
the performances of the two algorithms in a comparatively heavy workload scenario.  Furthermore, the
POD-TAS algorithm can be expanded to support computing chips with integrated GPU and CPU
cores by treating the GPU as another CPU core, to which only certain tasks (GPU
computation) may be assigned.

It is important to note that the POD-TAS algorithm differs from RT-TAS~\cite{lee_2019} in its use of a
dynamic thermal model. This change enables the prediction of how long a certain task can
execute on a CPU core before crossing the thermal threshold.  RT-TAS~\cite{lee_2019} relies on a
steady-state thermal model, which only predicts the temperature where the CPU will reach
thermal equilibrium for a certain power dissipation.  This allows RT-TAS~\cite{lee_2019} to use a task
assignment algorithm based on the Worst-Fit Decreasing (WFD) bin-packing algorithm to
maintain a balance among the steady state temperatures of the CPU cores. However, this
ignores the transient thermal behavior of the CPU cores caused by variations in the power
dissipation throughout task execution. The POD-TAS algorithm tracks the state of each CPU
core over time using a state transition diagram which further aids in assigning tasks to
suitable cores.  Thus, the POD-TAS algorithm ensures that the dynamic temperature of each
core does not exceed a given temperature threshold, which is the key to improving the
reliability and extending the lifespan of the processor chips. 

\section{Evaluation Methodology}
\label{sec:methods}

\begin{figure}[t]
    \centering
    \includegraphics[width=0.5\textwidth]{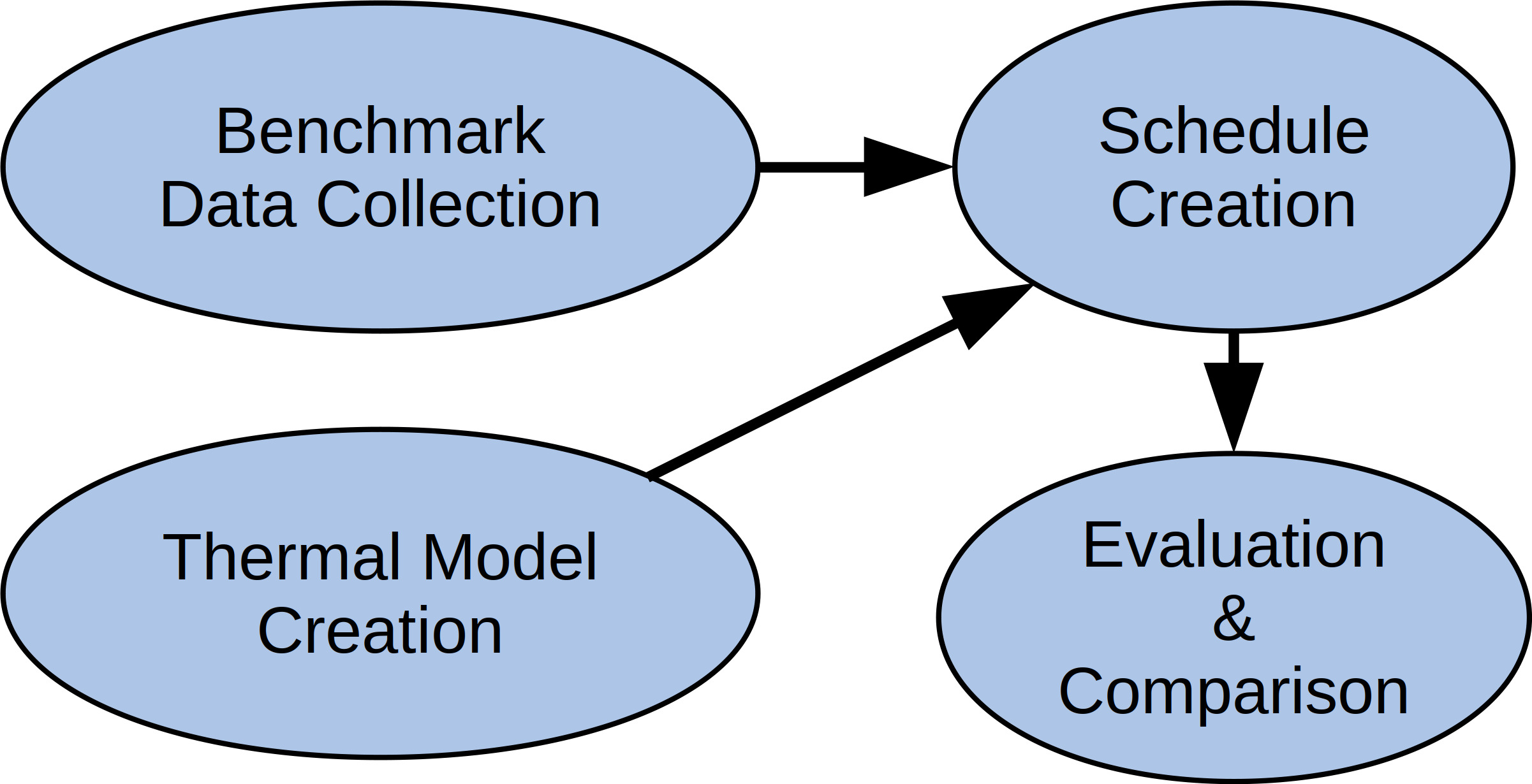}
    \caption{TAS Algorithm Evaluation Flow Overview}
    \label{fig:fullflow}
\end{figure}

\begin{figure}[t]
\begin{minipage}{\textwidth}
    \begin{minipage}{0.48\textwidth}
        \centering
        \includegraphics[width=0.9\textwidth]{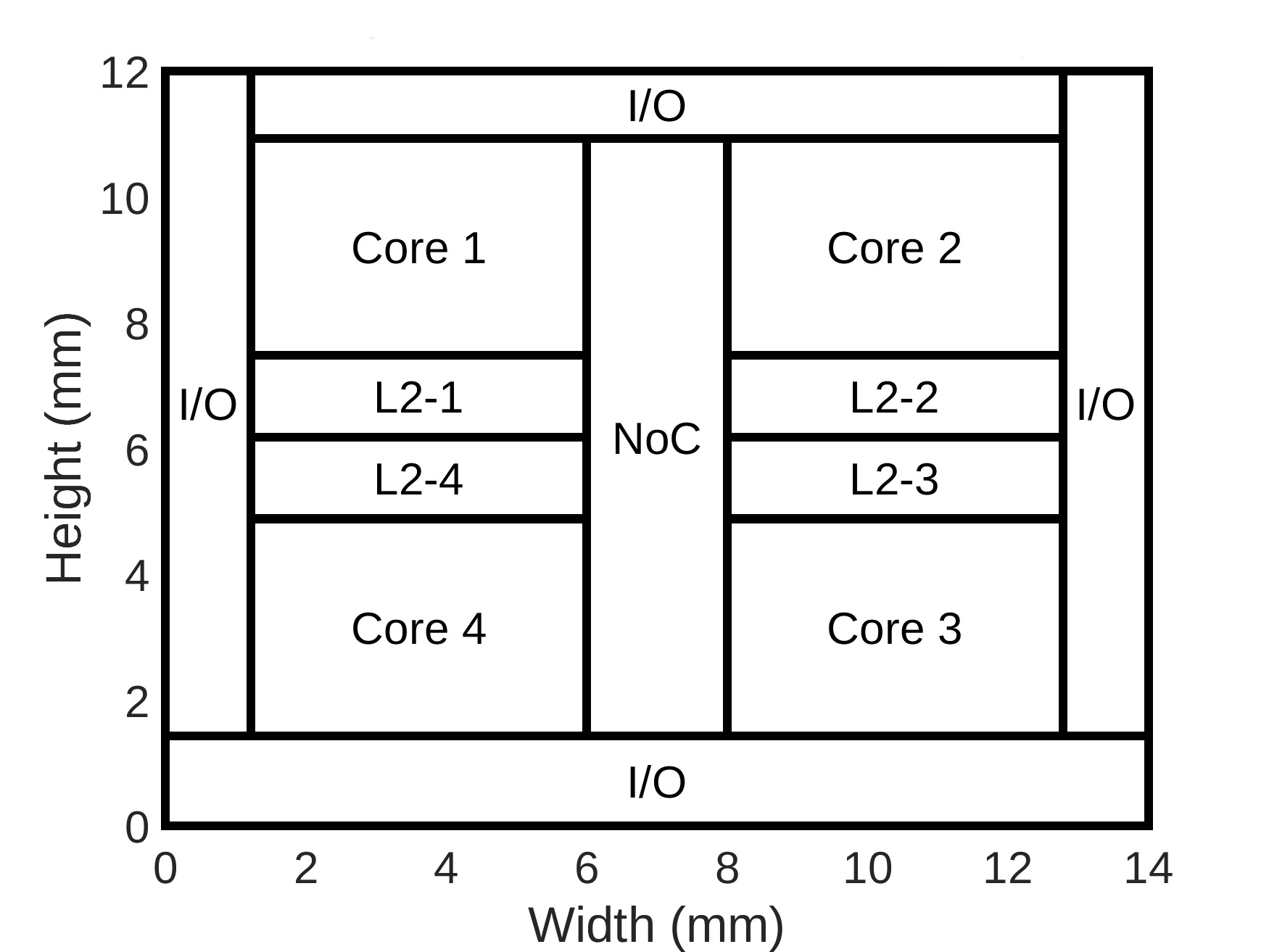}
        \captionof{figure}{Floorplan of AMD Athlon II X4 640}
        \label{fig:flp}
    \end{minipage}
    \begin{minipage}{0.48\textwidth}
        \centering
        \captionof{table}{CPU Configuration Parameters for AMD Athlon II X4 640}
        \begin{tabular}{|c|c|}
            \hline
            \textbf{Parameter} & \textbf{Value} \\ \hline\hline
            Clock Speed        & 3.0 GHz        \\ \hline
            Number of Cores    & 4              \\ \hline
            L1$_i$ Size        & 64 kB/core     \\ \hline
            L1$_d$ Size        & 64 kB/core     \\ \hline
            L2 Size            & 512 kB/core    \\ \hline
            Memory Size        & 8 GB           \\ \hline
            Memory Type        & DDR3 1600 8x8  \\ \hline
            L1 Cache Latency   & 3              \\ \hline
            L2 Cache Latency   & 15             \\ \hline
        \end{tabular}
        \label{tab:gem5conf}
    \end{minipage}

    \vspace{0.5em}
    \begin{minipage}{0.99\textwidth}
        \centering
        \captionof{table}{Selected Benchmarks from COMBS}
        \vspace{-0.5em}
        \label{tab:benches}
        \begin{tabular}{|c|c|c|}
            \hline
            \textbf{Benchmark}  & \textbf{Description} & \textbf{WCET (ms)}\\ \hline\hline
            2D Heat             & Tightly coupled heat distribution algorithm       & 147 \\ \hline
            Radix Sort          & Sorts a large array with the Radix Sort algorithm & 85 \\ \hline
            Advection-Diffusion & One dimensional CFD simulation                    & 41 \\ \hline
            Monte-Carlo         & Performs Monte-Carlo simulation to estimate $\pi$ & 32 \\
            \hline\hline
            FFTW              & Fastest Fourier Transform in the West               & 150 \\ \hline
            KS-PDE            & 1D Kuramoto-Sivashinsky equation                    & 84  \\ \hline
            Loops             & Speed-tests loop structures                         & 39 \\ \hline
            Lid-Driven Cavity & CFD simulation of viscous incompressible fluid flow & 78 \\ \hline
        \end{tabular}
    \end{minipage}
    \vspace{0.5em}
\end{minipage}
\end{figure}

With the goal of uniformly evaluating TAS algorithms with high spatiotemporal resolution,
we develop a simulation-based methodology. This methodology is used to evaluate both our
POD-TAS and the RT-TAS~\cite{lee_2019} algorithms using the same set of tasks (Table~\ref{tab:benches}) on the
same CPU platform.  To this end, a set of open-source simulators are adapted to be used
for evaluation.  Many steps are involved in the evaluation pipeline to enable a thorough
evaluation. First, the benchmark programs used as workloads for task scheduling are
evaluated to find their timing and power information. We also train the thermal models
that will be used for scheduling. Then, the trained thermal model and the task information
are given to the scheduler program, which constructs the execution schedule.  Then, this
schedule is executed in the final evaluation stages to gather the dynamic temperature
profile of the CPU during the execution of the schedule. An overview of the evaluation
process is shown in Figure~\ref{fig:fullflow}.

The simulators chosen are gem5~\cite{gem5} and Multicore Power, Area, and Timing
(McPAT)~\cite{mcpat_2009}.  Gem5 is a cycle-accurate CPU simulator that can be used to
gather traces of architecture-level events during program execution. These traces of event
statistics are then used as input for McPAT.  The McPAT simulator computes the power
dissipation of different FBs of the CPU based on the statistics trace from gem5. Lastly,
for thermal evaluation, the Finite Element Computational Software
(FEniCS)~\cite{fenics_2003} FEM platform is used to calculate the detailed thermal profile
of the CPU during execution.  The input to FEniCS is a floor plan of the CPU FBs, along
with the dynamic power map for each FB that was collected by McPAT.

The CPU chosen for our evaluation is the AMD Athlon II X4 640~\cite{amd_640}.  This CPU
model was chosen due to its quad-core configuration and the availability of information
about the layout of the CPU's FBs. The layout of the CPU is shown in Figure~\ref{fig:flp}.
Each FB of the CPU has a different purpose (e.g., northbridge, caches, cores, etc.).  Each
FB in the figure is labeled with its name.  For our evaluation, gem5 and McPAT are
configured to match the AMD Athlon II X4 640 as best as possible. The values used for the
CPU configuration are shown in Table~\ref{tab:gem5conf}. Thus, the simulators are matched
to the realistic CPU while achieving greater temporal resolution.

To provide a realistic workload for the CPU to run during thermal-aware task scheduling,
we select a subset of the open-source COMBS benchmark suite~\cite{combs_2020}.  This
benchmark suite was chosen to supply our workload because it contains many benchmarks that
are representative of realistic workloads, such as heat simulation,
Fourier transforms, PDE solving, and conjugate gradient calculation. 
The chosen subset of the COMBS benchmark suite used in our evaluation is described in
Table~\ref{tab:benches}.  This subset is chosen as the other benchmarks in the COMBS suite
exhibited compatibility issues with the Linux environment simulated by gem5 that is used
for our evaluation.  The evaluation has been performed using two different sets of the
benchmarks, namely, a set of 4 tasks and a set of 8 tasks.  We chose to evaluate with a
set of 4 tasks to create a scheduling scenario similar to the evaluation performed in Lee
et al.'s RT-TAS~\cite{lee_2019}.  The 4 task case provides a lighter scheduling scenario.
We evaluate an 8 task case where the thermal dissipation of the schedule is comparatively
higher than the 4 task case.  For the 4 task evaluations, the first 4 benchmarks in
Table~\ref{tab:benches} are used, while for the 8 task evaluations, all 8 of the listed
benchmarks are used.

\begin{figure}
    \centering
    \subcaptionbox{POD Model Creation\label{fig:podmodel}}{%
        \includegraphics[width=0.45\textwidth]{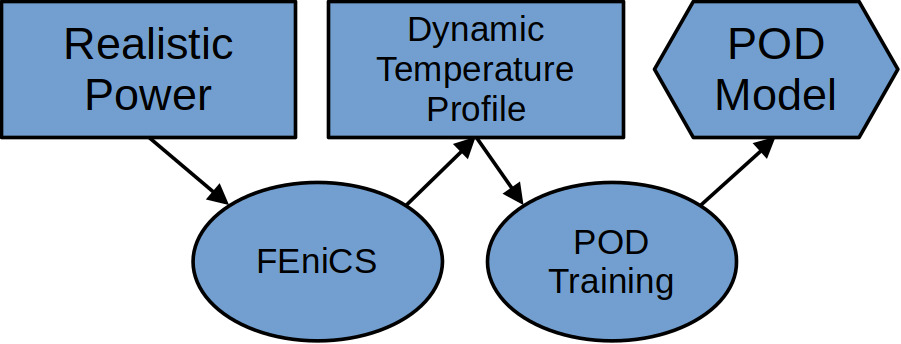}%
    }\hspace{0.05\textwidth}
    \subcaptionbox{Steady State model Creation\label{fig:ssmodel}}{%
        \includegraphics[width=0.45\textwidth]{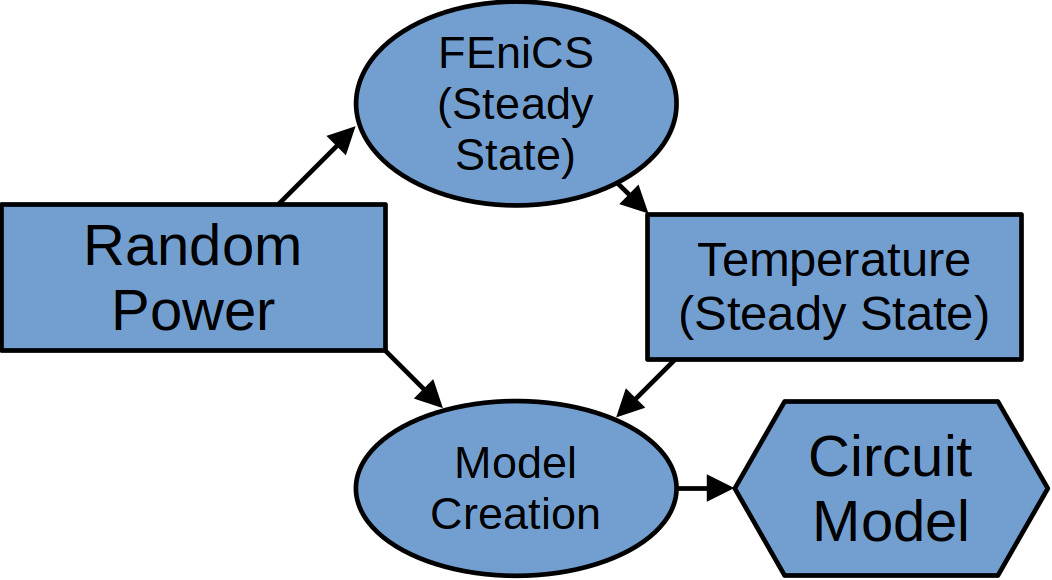}%
    }
    \caption{Thermal Model Creation for Use During Scheduling}
    \label{fig:modelcreation}
\end{figure}

Figure~\ref{fig:modelcreation} shows the flowcharts for the creation for both of the
thermal models (POD and steady-state) used in this work.  Figure~\ref{fig:podmodel}
shows the POD model creation flow. 
The temperature profile output by FEniCS is used in the POD training process described in
Eq.~\ref{eq:3} of Section~\ref{sec:pod}. This output creates a POD model that is trained
for the AMD Athlon II X4 640. 
For the creation of a steady-state model, we use the process
proposed by RT-TAS~\cite{lee_2019}. A flowchart of the process to
construct the steady-state model is shown in Figure~\ref{fig:ssmodel}.
To create the model, we first use FEniCS to find the steady-state temperature of the CPU
when a random power is applied to a single FB of the CPU. Then, from the temperature
solution output by FEniCS, we extract the average temperature of each FB to solve
for the thermal-coupling coefficients for each block needed for steady-state temperature
prediction. Through this, we train the steady-state model needed for prediction.
This thermal model is used by Lee et al. (2019) for
their scheduling algorithm, RT-TAS~\cite{lee_2019}.

\begin{figure}
    \centering
    \subcaptionbox{Benchmark Data Collection Flow\label{fig:benchdat}}{%
        \includegraphics[width=0.50\textwidth]{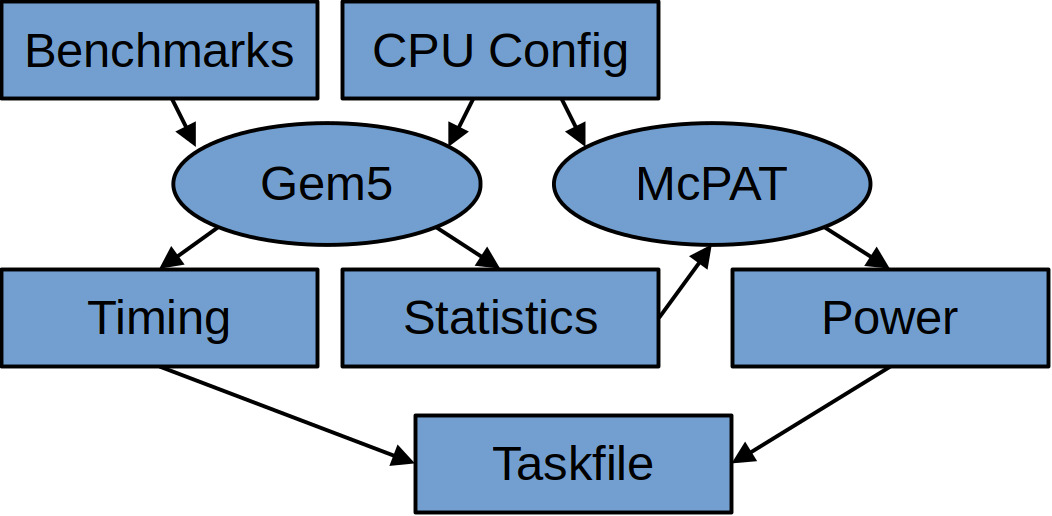}%
    }\hspace{0.04\textwidth}
    \subcaptionbox{Schedule Creation\label{fig:schedcreate}}{%
        \includegraphics[width=0.33\textwidth]{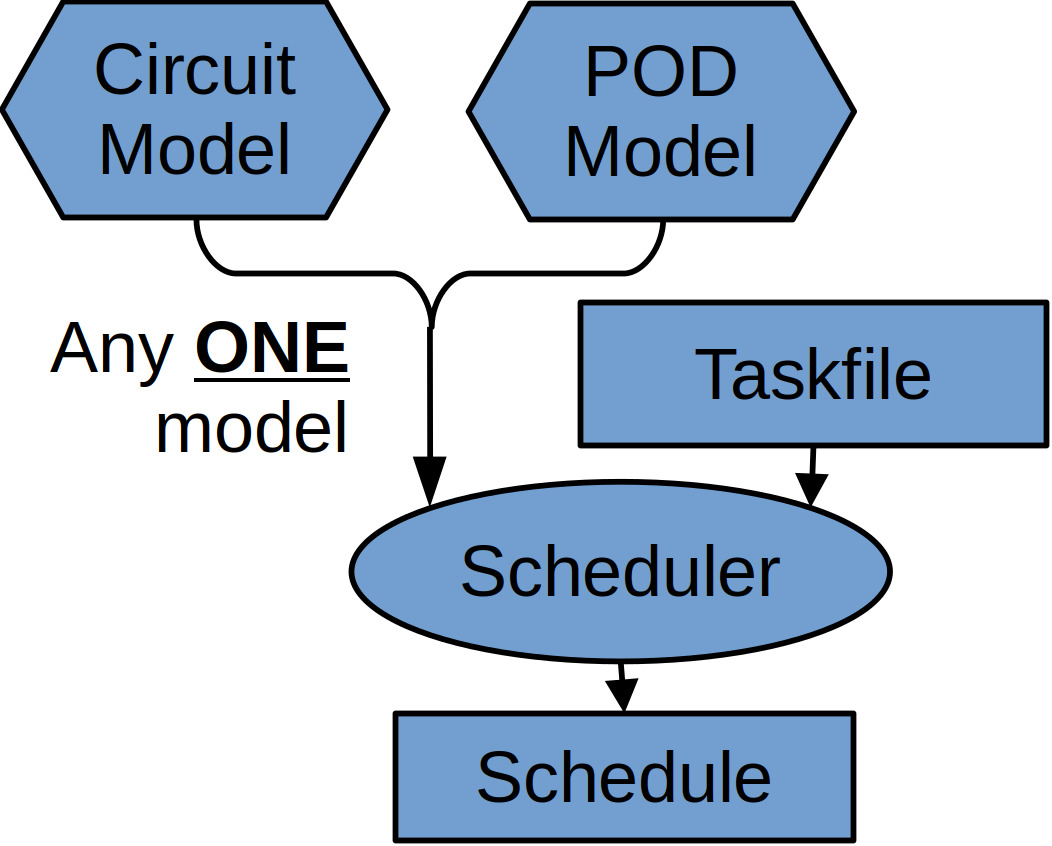}%
    }
    \caption{Benchmark Data Collection and Schedule Creation Flows}
    \label{fig:benchandsched}
\end{figure}

To supply the task scheduler with information about the
tasks, they are each first evaluated to collect the needed
information.  The timing information and power information
for each benchmark is used to create a task file to use
during scheduling. This task file includes information for
the scheduler about each task, including its execution time,
power dissipation, and name.  The flow of this process is
shown in Figure~\ref{fig:benchdat}.

The final stage before schedule evaluation is the creation of the execution
schedule. This process is shown in Figure~\ref{fig:schedcreate}. Using the task
file created during benchmark evaluation and one of the trained thermal models (either our POD or Lee et al.'s model~\cite{lee_2019}),
the scheduler is run to create the task execution schedule. If the thermal circuit model is
selected for evaluation, then the RT-TAS~\cite{lee_2019} algorithm is used. When the POD model is
chosen for evaluation, then the POD-TAS algorithm is used to create the schedule. This outputs
a schedule that can be used during the final evaluation.
Once schedules are created using both the thermal circuit model
from Lee et al. (2019)~\cite{lee_2019} and our own POD model,
they are evaluated to compare the thermal behavior of the
CPU during schedule execution.

\section{Evaluation Results}
\label{sec:eval}

\begin{figure}
    \begin{minipage}[b]{0.37\textwidth}
        \centering
        \includegraphics[width=\textwidth]{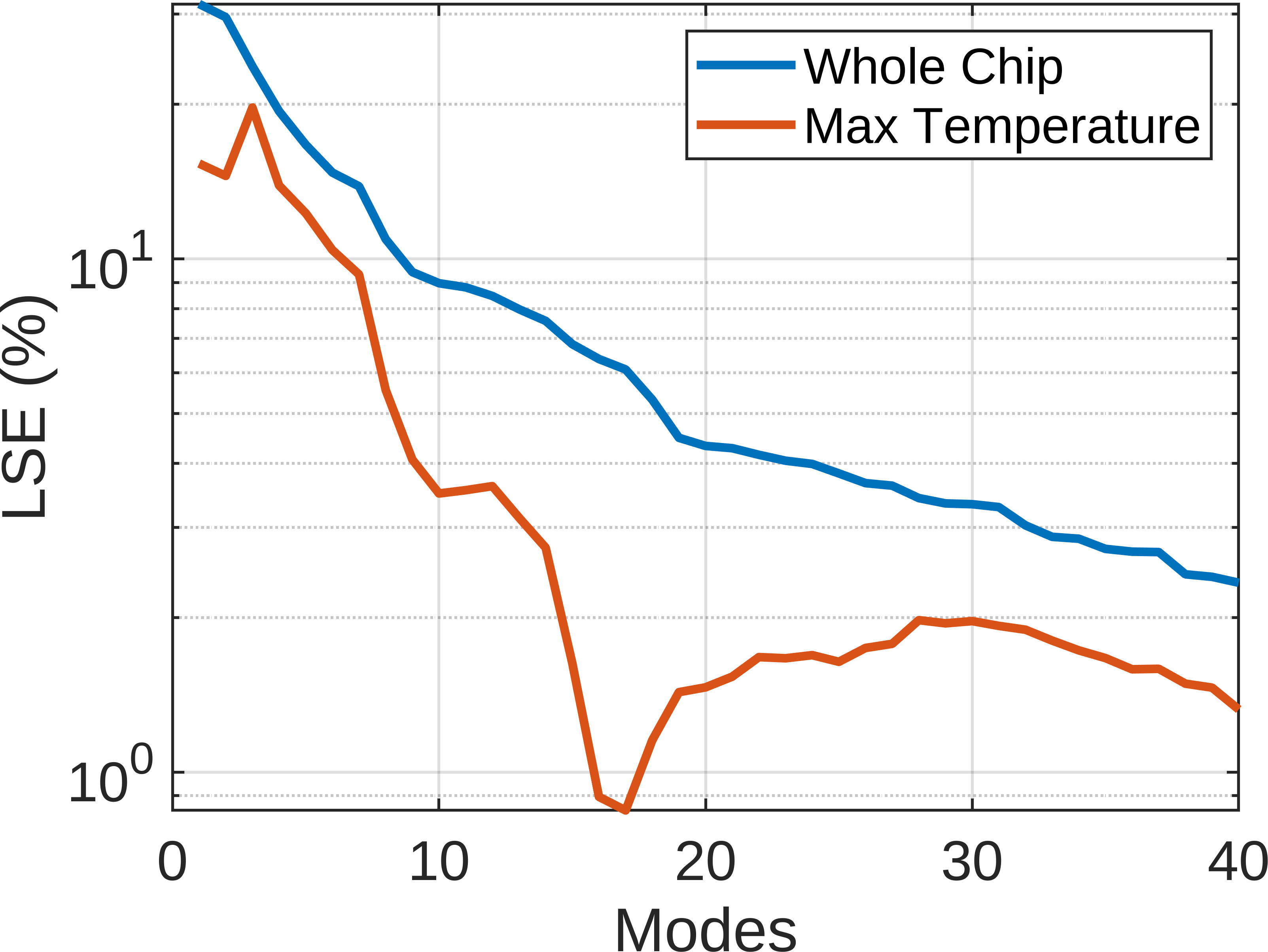}
        \captionof{figure}{Average LSE per number of POD Modes over 1 second of prediction}
        \label{fig:avg_lse}
    \end{minipage}\hfill%
    \begin{minipage}[b]{0.55\textwidth}
        \centering
        \includegraphics[width=0.8\textwidth]{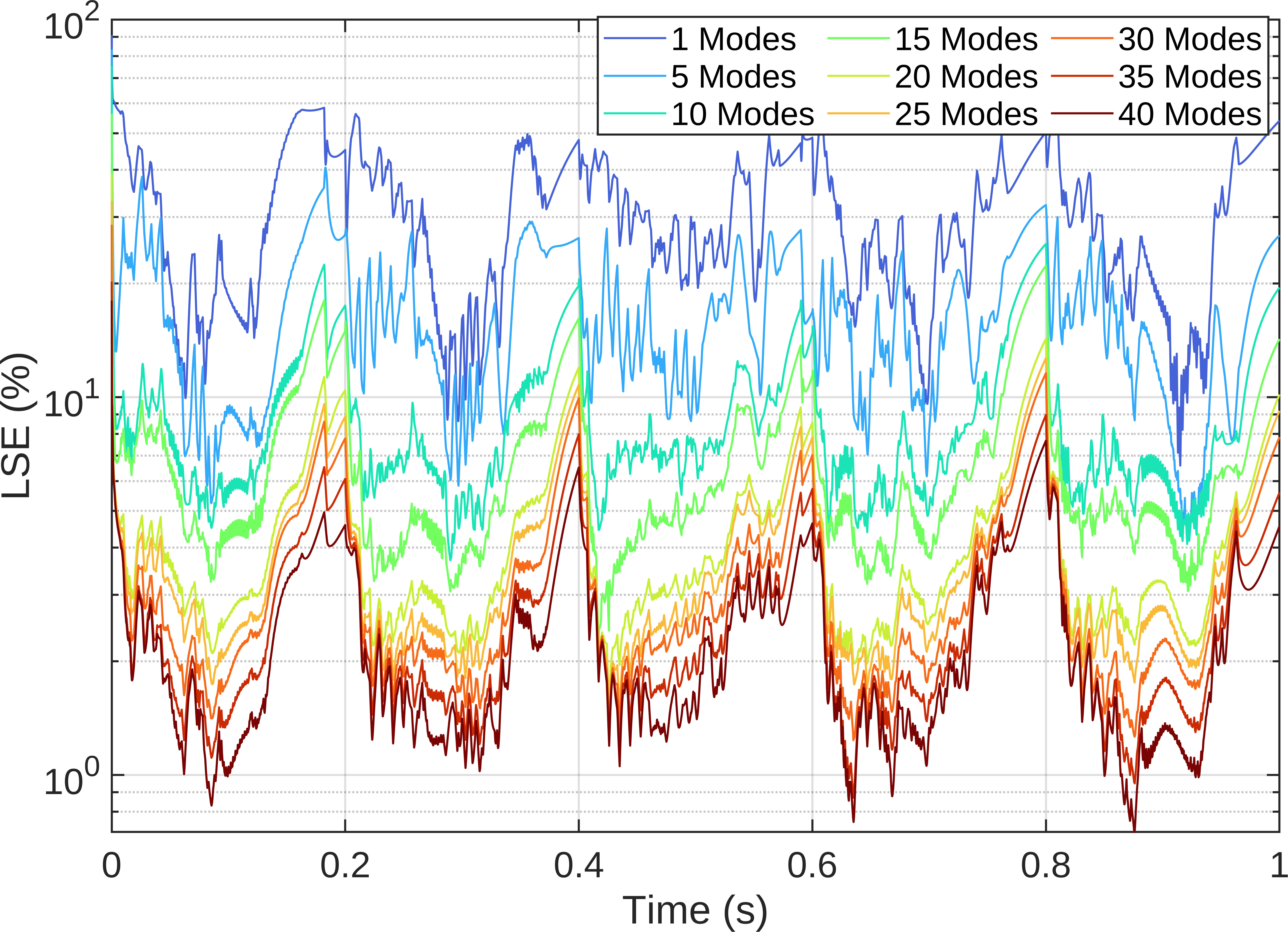}
        \captionof{figure}{Whole chip LSE over time per number of POD Modes}
        \label{fig:lse_time}
    \end{minipage}
    \vspace{0.5em}
\end{figure}

We perform evaluation using the methodology described in Section~\ref{sec:methods} on
both our POD-TAS algorithm and RT-TAS~\cite{lee_2019}.
To perform evaluation with our POD model, we first need to determine an
appropriate number of POD modes to use for prediction
during scheduling. To this end, a POD model is trained
for the CPU and is evaluated on a 
task schedule to simulate how it would perform when
predicting the temperature of the CPU during schedule
execution. To provide a ground-truth to compare the POD predictions
    to, FEniCS (FEM DNS) is used to calculate the temperature profile
    of the same schedule. Comparison is done using
    Least-Squared Error (LSE) as the metric. The LSE per time step is calculated using
Eq.~\eqref{eq:lse}, where $N_x$, $N_y$, and $N_z$ are
the number of thermal simulation mesh grid steps in the $x$, $y$, and $z$ directions
respectively; $T_i$ is the temperature at the
    $i^{\text{th}}$ point in space calculated by FEniCS (the ground-truth); and
$P_i$ is the temperature at the $i^{\text{th}}$ point in
space predicted by POD.

\begin{equation}
    \text{LSE}(\%) = 100*\sqrt{ \frac{\sum^{N_x*N_y*N_z}_{i=0}(T_i-P_i)^2}{\sum^{N_x*N_y*N_z}_{i=0}T_i^2}}\label{eq:lse}
\end{equation}
This gives
prediction results with LSEs over time
shown in
Figure~\ref{fig:lse_time}.  A key behavior of these
error measurements is that, while they vary over time, they
do not have an increasing trend. This shows
that the predictions of the POD model remain accurate over a
very long period of time. The averages over time of the LSE per
number of modes are shown by the blue line in
Figure~\ref{fig:avg_lse}. The red line in
Figure~\ref{fig:avg_lse} shows the LSE of the maximum
predicted temperature in the active layer of the chip. This
is a key value because we are most interested in the peak
temperature of the chip during scheduling. Based on
Figures~\ref{fig:avg_lse} and~\ref{fig:lse_time}, the
predictions for schedule construction in POD-TAS for the
remainder of this work are performed using 30 modes
to balance the accuracy of the model and computational
overhead. Furthermore, based on
Figure~\ref{fig:podhotfem}, we can see that 30 modes
    follows the temperature calculated with FEniCS (FEM) very well.

\begin{figure}
    \centering
    \subcaptionbox{4 Tasks\label{fig:4taskscheds}}{%
        \includegraphics[width=0.48\textwidth]{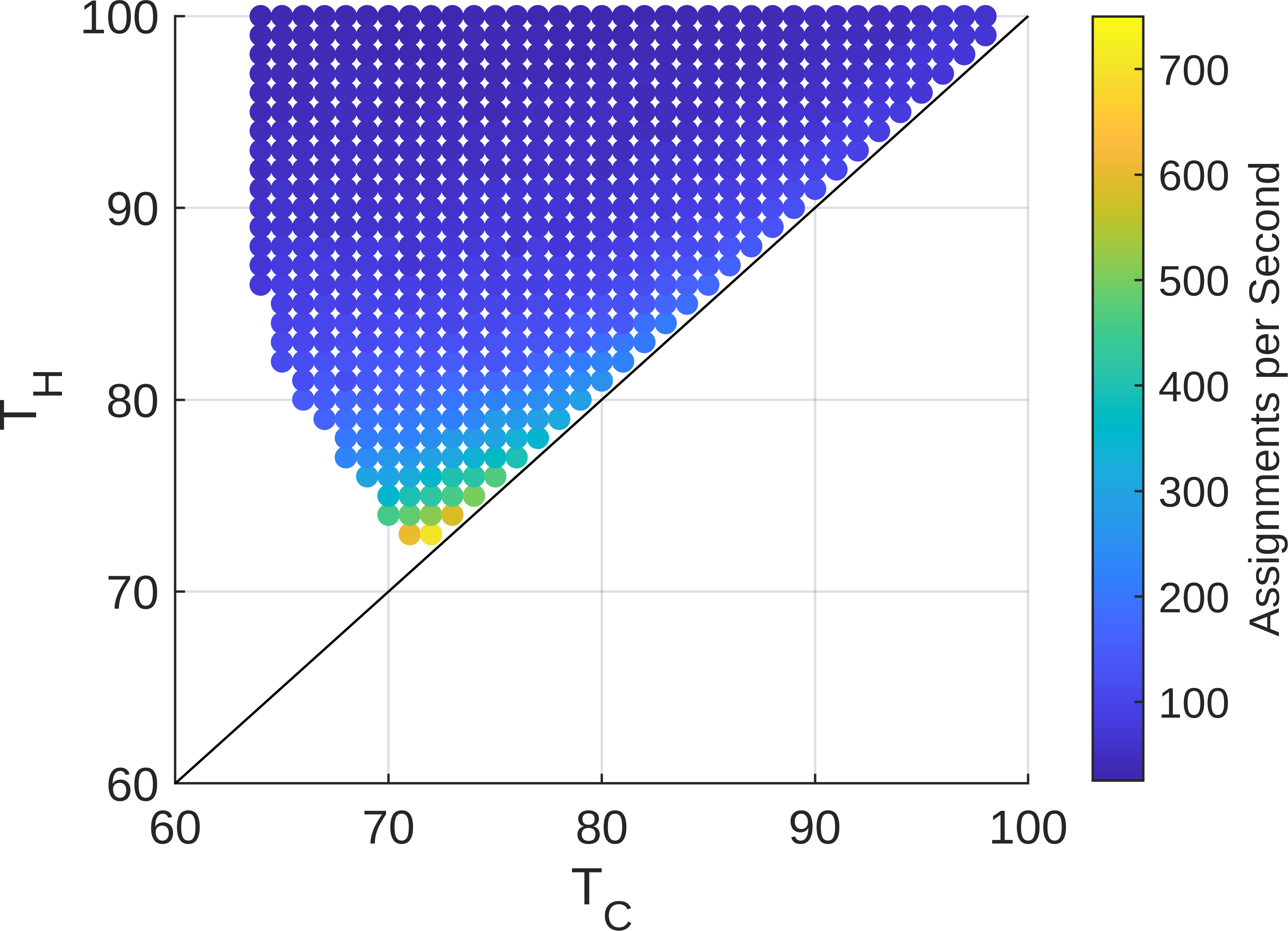}%
    }\hspace{0.02\textwidth}
    \subcaptionbox{8 Tasks\label{fig:8taskscheds}}{%
        \includegraphics[width=0.48\textwidth]{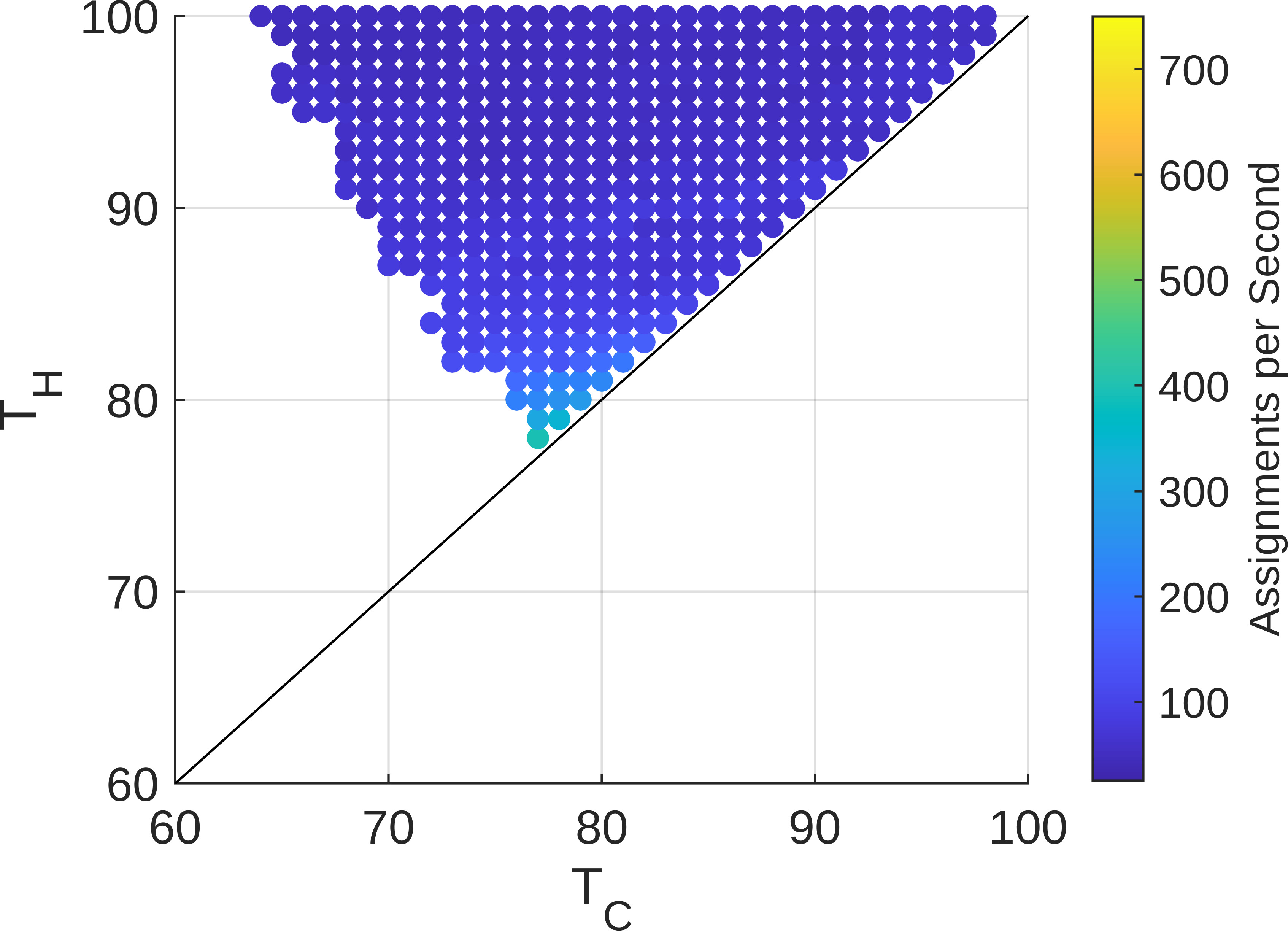}%
    }
    \caption{POD-TAS task assignment frequency vs $T_H$ and $T_C$}
    \label{fig:scheds}
\end{figure}

Figure~\ref{fig:scheds} shows some characteristics of the
schedules for varying combinations of $T_C$ and $T_H$.
Figure~\ref{fig:4taskscheds} and
Figure~\ref{fig:8taskscheds} show the task assignment
frequency for the 4 task evaluation set and the 8 task
evaluation set respectively.  Note that pairs of $(T_C,T_H)$
where $T_C\geq T_H$ are invalid; thus, there are no points
shown on or below the line $T_H=T_C$. Other points that are
missing in the plot are due to those combinations of
$(T_C,T_H)$ not yielding a schedule that meets the hard
real-time deadlines for all tasks. By comparing
Figure~\ref{fig:4taskscheds} and
Figure~\ref{fig:8taskscheds}, we can observe that as more
tasks are added without changing the deadlines of any tasks,
the values of $T_H$ and $T_C$ must be increased to create
valid schedules. Furthermore, as $T_H$ approaches $T_C$ or
as $T_H$ is lowered, the frequency of assignments performed
by POD-TAS increases. This is due to a requirement for the
algorithm to more actively control the thermal dissipation
of the chip to maintain the temperature within the
constraints.

\begin{figure}[t]
    \centering
    \includegraphics[width=0.95\textwidth]{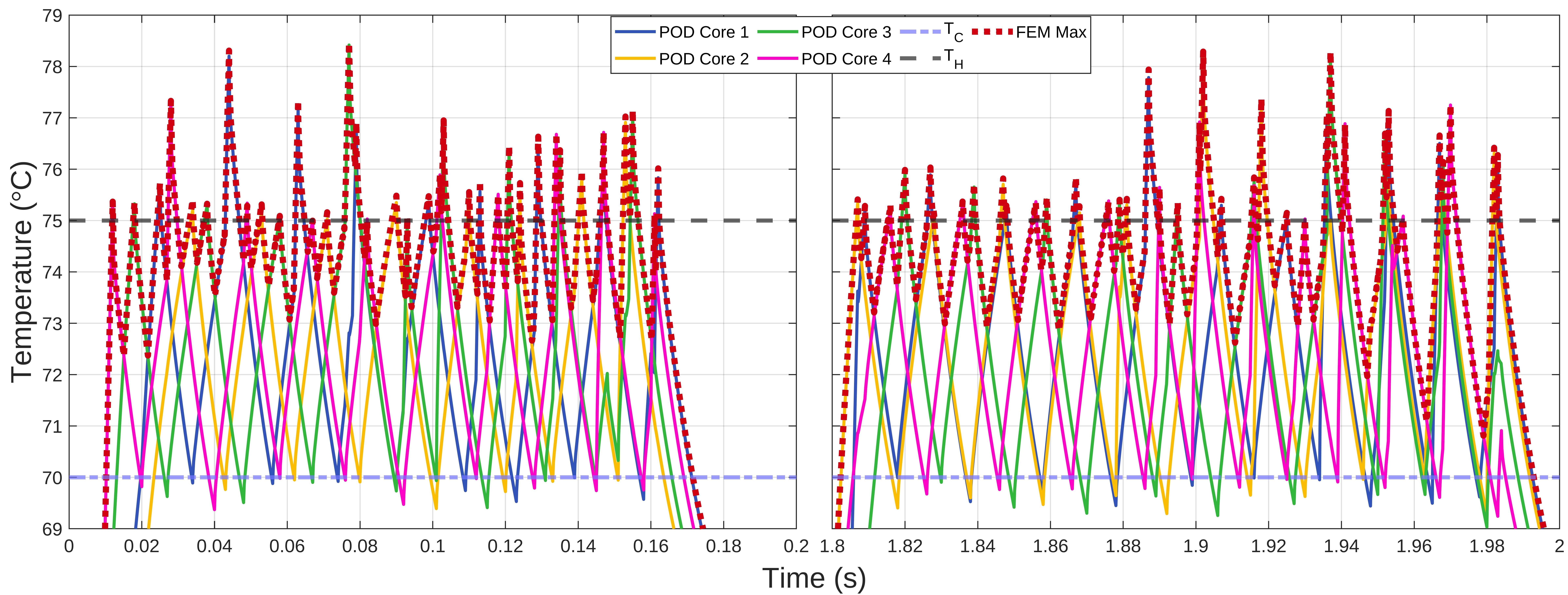}
    \caption{POD predictions during schedule construction}
    \label{fig:pod_pred}
\end{figure}

Figure~\ref{fig:pod_pred} shows an example of the
temperature of the CPU as calculated with FEniCS during the
execution of the schedule compared to the temperatures of
the CPU that POD predicted during scheduling. For this case,
$T_H=75^{\circ}$C and $T_C=70^{\circ}$C.  The left-hand side of
Figure~\ref{fig:pod_pred} shows the first 200ms of schedule
construction, while the right-hand side shows the predictions
between 1800ms and 2s.  The POD predictions follow the
maximum temperature from FEniCS very closely. Furthermore, we
can see that each core follows the behavior defined by the
CPU states shown in Figure~\ref{fig:cpustates} after
heating up beyond $T_H$. The cores that have exceeded $T_H$ idle to cool down until its
temperature is below $T_C$.

\begin{figure}[t]
    \centering
    \includegraphics[width=0.95\textwidth]{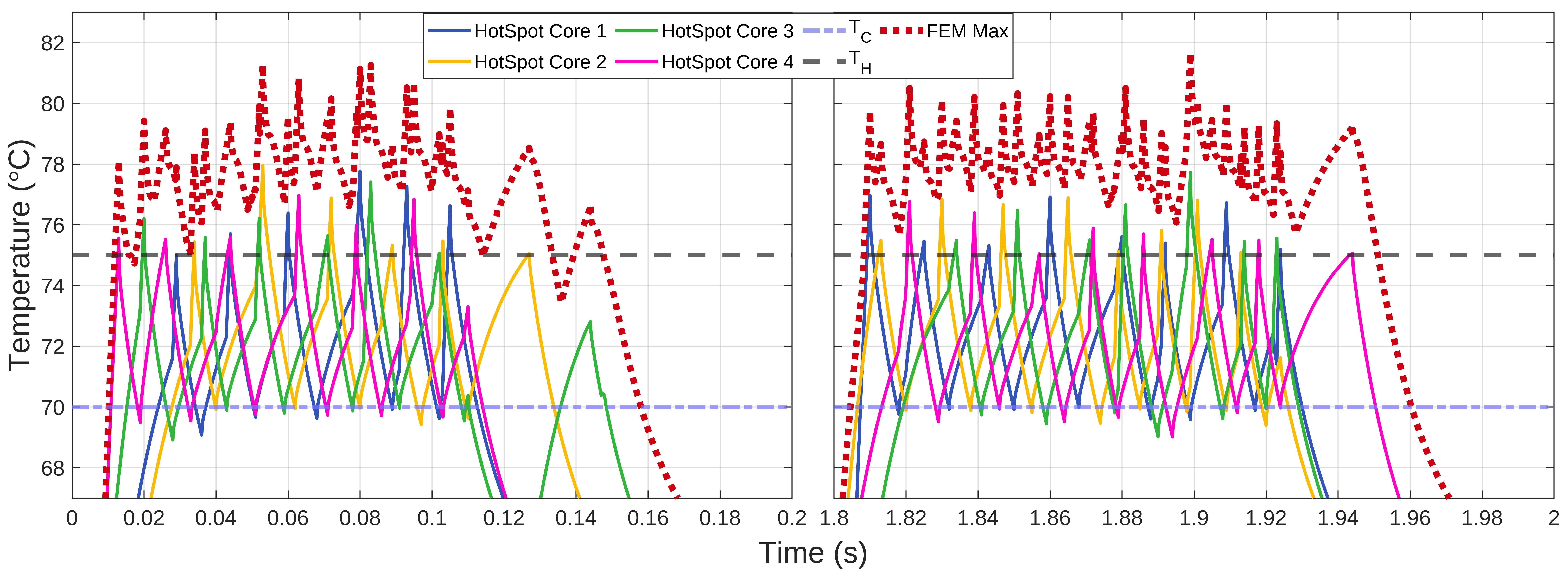}
    \caption{HotSpot predictions during schedule construction}
    \label{fig:hs_pred}
\end{figure}

Figure~\ref{fig:hs_pred} shows the result of using
HotSpot as the thermal prediction model during scheduling
with the POD-TAS algorithm. This is performed using the same
CPU and task configuration as the four task POD result shown
in Figure~\ref{fig:pod_pred}. HotSpot is configured
with a mesh size of $256\times256\times17$ to match the
POD model as closely as possible. This demonstrates the need
for an accurate thermal model during scheduling. We can see
from Figure~\ref{fig:pod_pred} that the POD predictions
are highly accurate, but when HotSpot is used, there is an
error of about 3.5$^\circ$C consistently.  This
corresponds to the temperature prediction above the ambient
temperature having an average percent error of 12.69\%. 
Furthermore, HotSpot tends to predict the
temperature to be below what the true temperature of the CPU
is while executing the schedule, leading to a much higher
temperature than is predicted during scheduling.
POD
demonstrates greatly reduced prediction error. This allows
POD to make scheduling decisions that reflect the true CPU
thermal state during execution.

\begin{figure}[t]
    \centering
    \subcaptionbox{POD-TAS 4 Task\label{fig:podtas4task}}{%
        \includegraphics[width=0.48\textwidth]{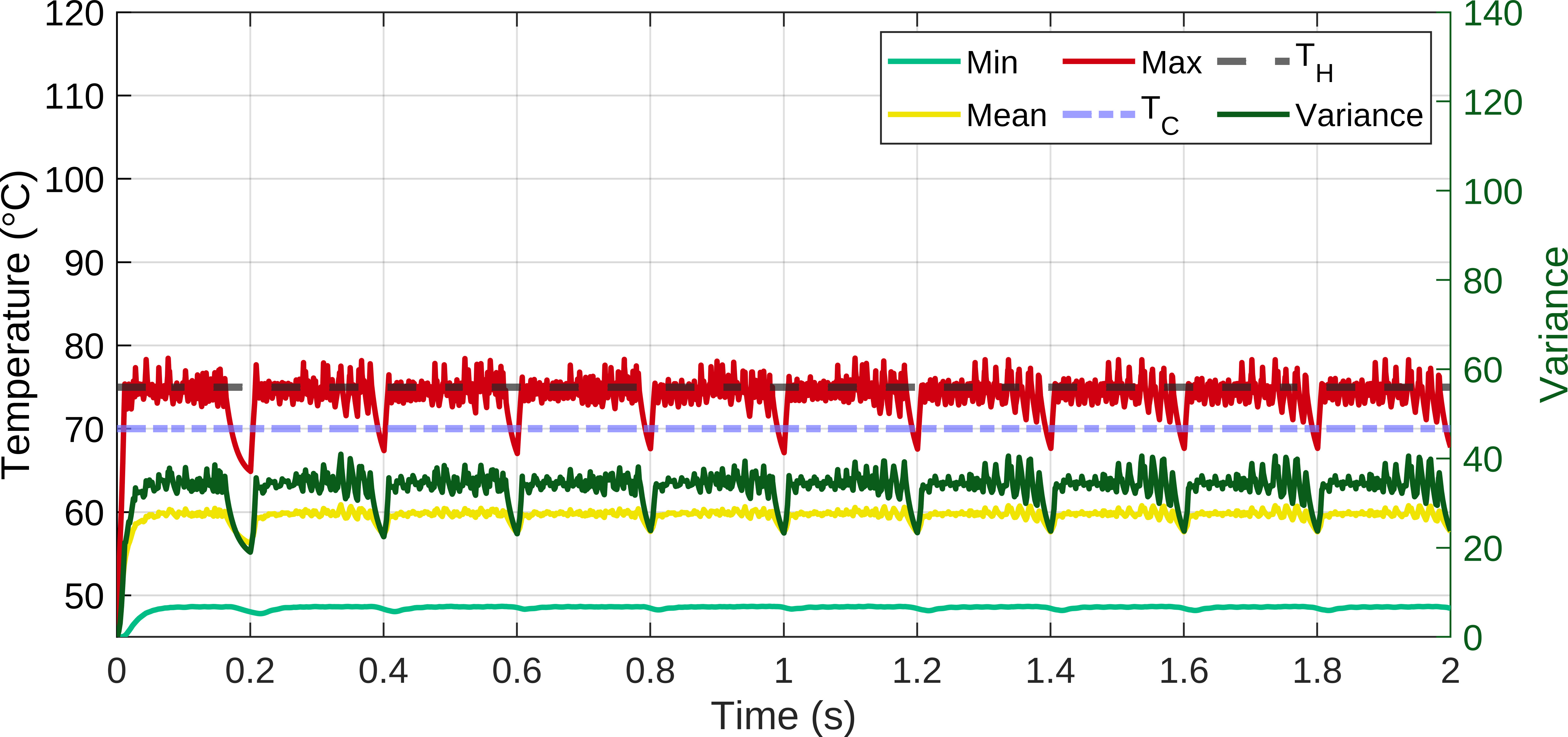}%
    }\hspace{0.02\textwidth}
    \subcaptionbox{RT-TAS~\cite{lee_2019} 4 Task\label{fig:rttas4task}}{%
        \includegraphics[width=0.48\textwidth]{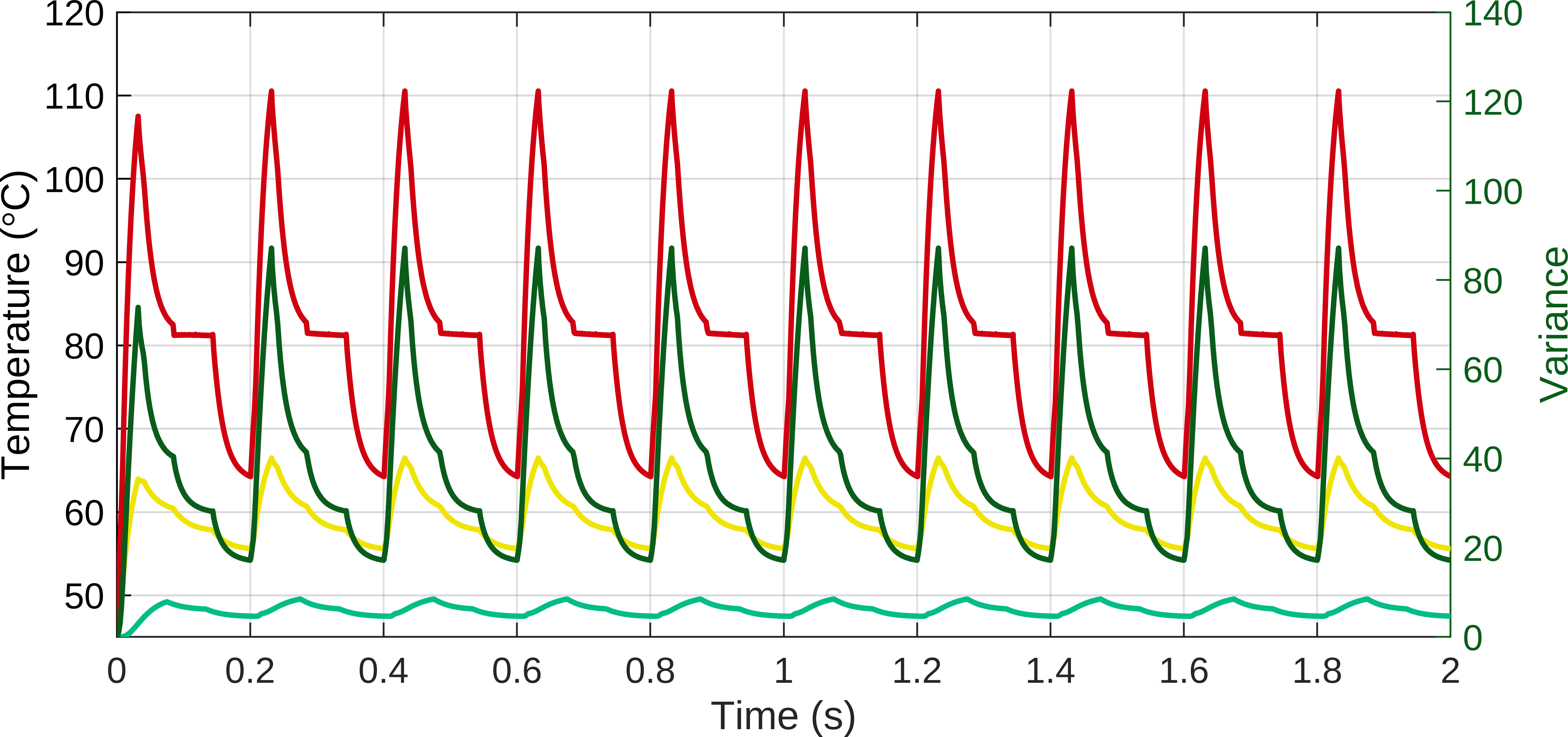}%
    }
    \subcaptionbox{POD-TAS 8 Task\label{fig:podtas8task}}{%
        \includegraphics[width=0.48\textwidth]{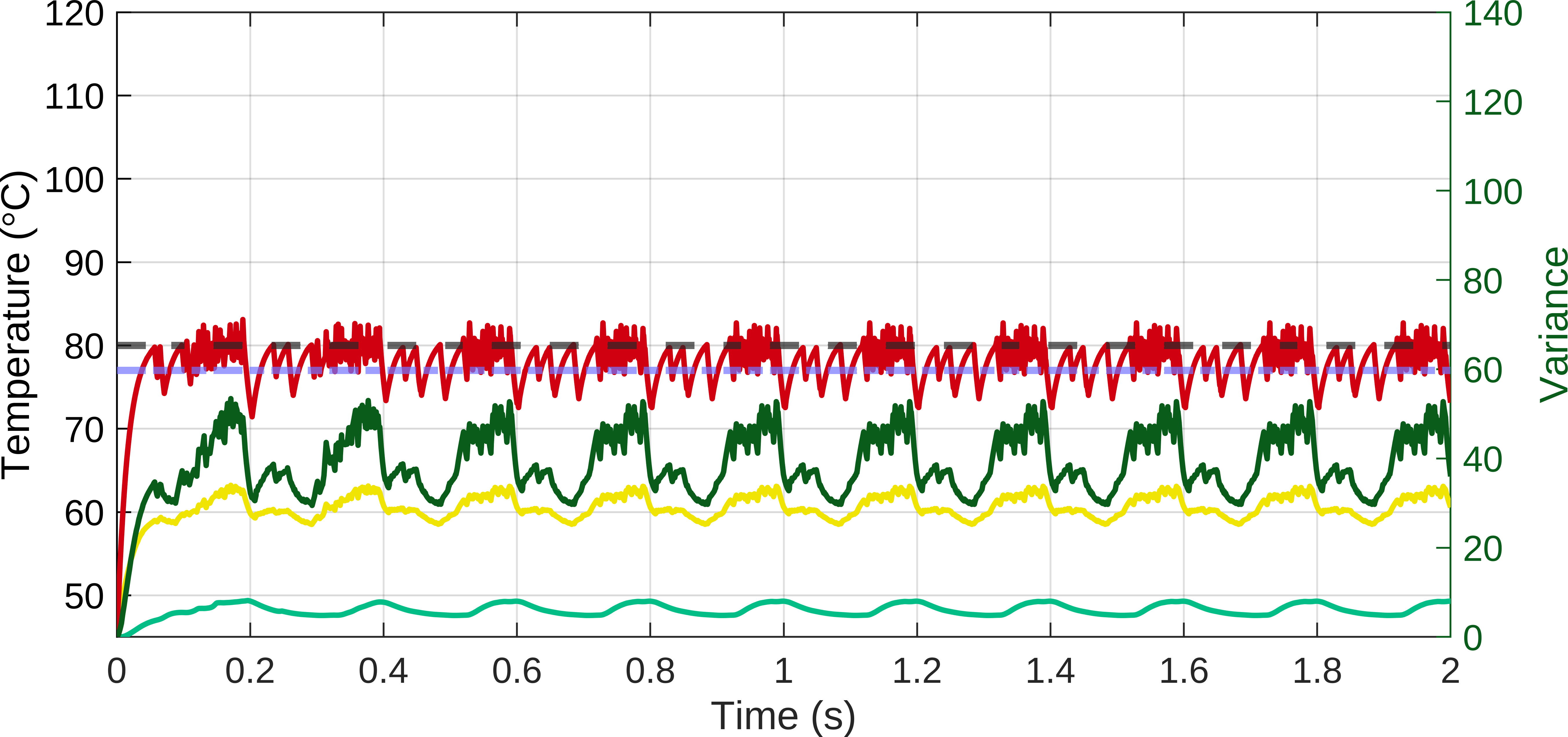}%
    }\hspace{0.02\textwidth}
    \subcaptionbox{RT-TAS~\cite{lee_2019} 8 Task\label{fig:rttas8task}}{%
        \includegraphics[width=0.48\textwidth]{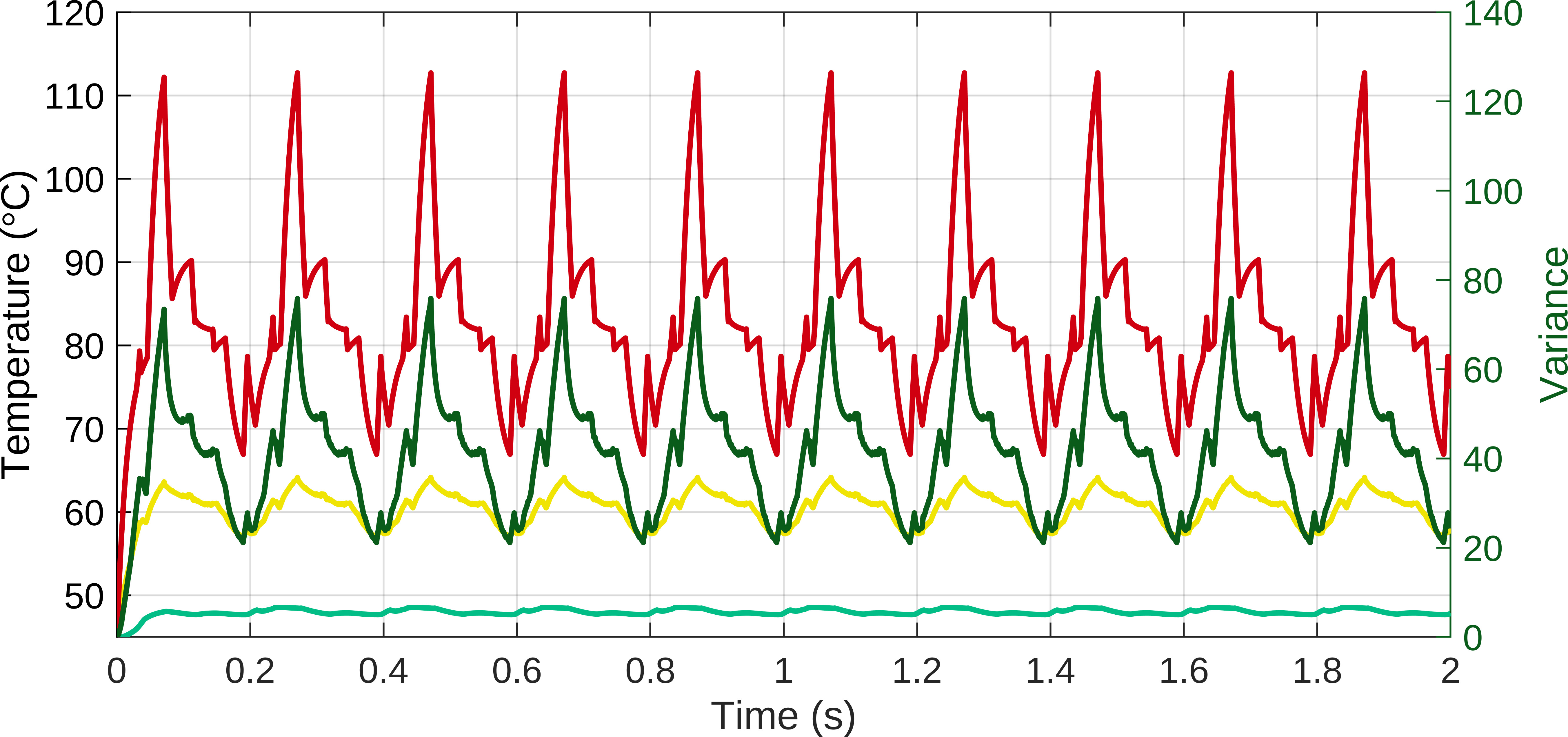}%
    }
    \caption{CPU temperature during schedule execution}
    \label{fig:results}
\end{figure}

To demonstrate the effectiveness of the POD-TAS algorithm,
we have evaluated it alongside the RT-TAS~\cite{lee_2019}
algorithm in our simulation platform.  The
evaluation for both algorithms is performed for a period of
2 seconds. The simulation uses a $\Delta t$ of 10$\mu$s to
provide a very high temporal resolution to view the
transient thermal behavior of the CPU. This high temporal resolution
allows for the capture of temperature spikes that occur in very
small time scales.
The thermal results
of this evaluation for both models are shown in Figure~\ref{fig:results}.
The left y-axis of each plot is the temperature, against
which the minimum, mean, and
maximum chip temperatures; and temperature thresholds ($T_C$,
$T_H$) are plotted. The right y-axis of each plot is the
thermal variance, against which the variance values are
plotted. The schedule for the evaluation of POD-TAS with 4
tasks is generated using $T_H=75^\circ\text{C}$ and
$T_C=70^\circ\text{C}$.
The 8 task case for POD-TAS uses thresholds of
$T_H=80^\circ\text{C}$ and $T_C=77^\circ\text{C}$. These thresholds are shown with
horizontal lines in Figures~\ref{fig:podtas4task} and~\ref{fig:podtas8task}. These
thresholds are selected based on the frequency of
assignments for the thresholds shown in
Figure~\ref{fig:scheds}. The highest frequencies of
assignments are avoided while aiming to minimize the value
of $T_H$. On the other hand, for our evaluation of RT-TAS~\cite{lee_2019},
shown in Figures~\ref{fig:rttas4task}
and~\ref{fig:rttas8task}, no values for any temperature thresholds are shown since RT-TAS~\cite{lee_2019} does not
support their use.

\newcommand{\rttas}{\textbf{RT-TAS~\cite{lee_2019}}}
\newcommand{\podtas}{\textbf{POD-TAS}}
\newcommand{\pchange}{\textbf{\% Difference}}

\begin{table}[t]
    \centering
    \caption{Evaluation Metrics to Compare the Performances of the POD-TAS and RT-TAS~\cite{lee_2019} Algorithms. All temperatures from FEniCS are calculated in $^\circ$C.}
    \label{tab:diffs}
    \begin{tabular}{|c||c|c|c||c|c|c|}
        \hline
                       &\multicolumn{3}{c||}{\textbf{4 Task}}&\multicolumn{3}{c|}{\textbf{8 Task}} \\ \hline
        \textbf{Metric}& \rttas & \podtas &\pchange          & \rttas & \podtas & \pchange \\
        \hline\hline
        Peak Temp. & 110.53 &   78.47 & -29.01           & 112.71 &   83.11 & -26.26  \\ \hline
        Peak Var.  &  87.12 &   40.95 & -53.00           &  75.83 &   53.41 & -29.57  \\ \hline
        Var(Mean)  &  10.08 &    1.14 & -88.69           &   5.04 &    3.03 & -39.88  \\ \hline
        Var(Max)   & 141.71 &    5.41 & -96.18           & 111.49 &    7.51 & -93.26  \\ \hline
        Var(Var)   & 332.80 &   15.03 & -95.48           & 178.30 &   53.28 & -70.12  \\ \hline
    \end{tabular}
\end{table}

Table~\ref{tab:diffs} gives more details regarding the
performance of the algorithms.  For evaluating both of the
algorithms, we utilize several performance metrics.  The
first metric is the peak chip temperature during schedule
execution. The second is the peak spatial variance of the
chip temperature. This metric is affected by the presence of
high thermal gradients within the chip. Lower thermal
gradients yield a lower spatial variance of the chip
temperature.  Variance is calculated as
$\sum_{i=1}^{n}(x_i-\bar{x})^2/n$, where $n$ is the number
of samples, $\bar{x}$ is the sample mean, and $x_i$ is the
$i$-th sample. The last three variance metrics in
Table~\ref{tab:diffs} are all calculated in a similar manner
to one another.  First, the given metric is calculated from
the spatial temperature distribution, for instance, the mean
temperature in the chip at each time step. Each time step of
the evaluation yields a single value for the chip mean, and
maximum temperature, and a single value of the variance of
the temperature in space. These values are plotted
in Figure~\ref{fig:results}. Then, to observe how steady each measurement
is, the variance of each metric is taken over time and
reported in Table~\ref{tab:diffs}.

From Table~\ref{tab:diffs}, we can see that POD-TAS strongly
outperforms RT-TAS~\cite{lee_2019} for both the 4 task and 8 task scenarios.
The peak chip temperature is lowered by 29.01\% in the 4 task
case and 26.26\% in the 8 task case by POD-TAS due to its
capability to maintain the temperature very close to the
configured thresholds. $T_H$ is exceeded in some cases, but
this is due to the temporal fidelity of the decision making
of the TAS algorithm, which is limited to only be able to
make decisions with a time difference of 1 millisecond.
However, we can see that the magnitude of exceeding $T_H$ is
very limited.  Furthermore, the peak spatial variance of the
temperature is reduced by 53.00\% and 29.57\% in the 4 and 8 task
cases respectively by POD-TAS.  This
demonstrates the ability of POD-TAS to restrict the occurence of
high thermal gradients in space.

Also displayed in Table~\ref{tab:diffs} is the percent
difference of each metric. The percent difference is
calculated as $100*(a-b)/b$ where $a$ is the value of the
metric when POD-TAS is used, and $b$ is the value when
RT-TAS~\cite{lee_2019} is used.  These values demonstrate that POD-TAS
maintains a much more stable chip temperature over time.
Especially important are the reductions in the variance of
the maximum chip temperature over time by 96.18\% and
93.26\% in the 4 and 8 task cases respectively.
Furthermore, the variance of the mean chip temperature
is reduced by 88.69\% in the 4 task case and 39.88\% in the
8 task case and the variance of the spatial variance is
reduced by 95.48\% and 70.12\% in the 4 and 8 task cases
respectively. These reductions in the metrics demonstrate
that the maximum chip temperature is maintained at a much
more consistent level by POD-TAS than by RT-TAS~\cite{lee_2019}. This effect
avoids repeated heating and cooling of the chip which helps
to improve chip reliability and prevent premature device aging. 

\section{Future Research Directions}
\label{sec:future}

Future work in this new direction can include a more
thorough investigation into the behavior of POD-TAS as the
parameters change. A greater understanding of how the
generated schedules change depending on the temperature
thresholds would help to more optimally apply POD-TAS.
Furthermore, evaluation with more varied task sets would
further deepen the understanding of the algorithm's
behavior. An expansion of this algorithm to be capable of
scheduling a sporadic task set and reacting to sudden
changes to the task set would further leverage the
capabilities of the underlying POD thermal model. This would
achieve the creation of an even more versatile algorithm
that can be applied to manage the thermal behavior of
systems with non-periodic tasks, such as laptops, desktops,
and server computers. This would also allow POD-TAS to be
applied in data center environments to optimize the thermal
dissipation of large-scale computing systems. Furthermore,
expanding the POD-TAS algorithm to perform co-scheduling on
computing platforms with integrated GPU and CPU cores would
enable its use in more diverse computing environments given
our implemented POD-TAS algorithm can be readily extended to
support this capability. Further expansion to the POD-TAS
algorithm would be the addition of a method to enable
the dynamic modification of the thresholds to minimize the
temperature dissipation under a sporadic workload.

\section{Conclusion}
\label{sec:conc}

TAS algorithms can employ a variety of methods to limit the thermal dissipation of a chip. These methods
include careful scheduling of task execution, the use of DVFS, and many others. Many TAS methodologies rely on
the use of a thermal model to predict the chip temperature during schedule execution.  These thermal models
can have limitations in prediction accuracy or efficiency that can negatively impact the algorithm's ability
to do effective task scheduling. In this work, we have proposed and implemented POD-TAS, a TAS algorithm that
is based on a reduced-order thermal model enabled by POD.

The POD thermal model enables the prediction of temperature with high spatiotemporal resolution.  The accuracy
of the dynamic thermal RC circuit simulator, HotSpot, is compared with the POD-based thermal model in
Figure~\ref{fig:podhotfem}. This comparison supports the use of the POD-based model for scheduling. The
dynamic nature of the POD thermal model enables POD-TAS to utilize a pair of temperature thresholds to manage
the transient CPU temperature during task scheduling. On the other hand, a steady state thermal model does not provide the
transient temperature predictions needed to control the chip temperature based on threshold values.  The
states of the CPU cores are defined using a state transition diagram (Figure~\ref{fig:cpustates}) that enables
POD-TAS to avoid the assignment of tasks to cores that are in unsuitable states.

We replicate the steady state thermal model by Lee et al.~\cite{lee_2019} and their TAS algorithm, RT-TAS~\cite{lee_2019}.
The RT-TAS~\cite{lee_2019} algorithm is evaluated alongside the implemented POD-TAS algorithm in a novel simulation-based
evaluation pipeline so that the performance of the two algorithms can be directly compared. Our evaluation
methodology captures the thermal behavior of the TAS algorithms being evaluated with high spatiotemporal
resolution over a period of 2 seconds. This enables sharp temperature variations to be observed. Our evaluation includes two subsets of
the COMBS~\cite{combs_2020} benchmark suite to provide a 4 task evaluation case similar to the evaluation case
used by Lee et al.~\cite{lee_2019} and a heavier 8 task case.  The evaluation of the POD-TAS algorithm
demonstrates its capabilities to control the CPU temperature based on a pair of configured temperature
thresholds.  Utilizing our novel evaluation methodology on both algorithms, it is observed that POD-TAS
consistently outperforms RT-TAS~\cite{lee_2019}.  Through our evaluation, it is observed that POD-TAS can reduce the peak
spatial variance of the CPU by 53\% and 29.57\%, and the peak chip temperature by 29.01\% and 26.26\% compared
to RT-TAS~\cite{lee_2019} in the 4 and 8 task evaluation cases respectively.  Furthermore, variations in chip temperature over
time are greatly reduced by POD-TAS. These improvements can help to greatly stabilize and reduce the
temperature over the entire chip and consequently improve the lifetime of a chip.

\section*{Acknowledgements}
This work is supported by National Science Foundation under
Grant No. ECCS-2003307.

\bibliographystyle{plain}
\bibliography{biblio}

\end{document}